\documentclass[twocolumn]{aastex63}

\usepackage{newtxtext,newtxmath}

\usepackage[T1]{fontenc}  \usepackage{ae,aecompl}
\usepackage{graphicx} \usepackage{amsmath}     \usepackage{amssymb}
\usepackage{epsfig}  \usepackage{lineno}  \usepackage{overpic}
\usepackage{hyperref}

\newcommand{\nuvr}{NUV$-r$} \newcommand{\mstar}{$M_\ast$}
\newcommand{\msolar}{M$_\odot$} \newcommand{\Lsolar}{L$_\odot$}
\newcommand{\hh}{H$_{2}$}

\newcommand{\tx}[1]{\textrm{#1}} \newcommand{\kms}{km~$\tx{s}^{-1}$}
\newcommand{\TA}{$T^{*}_{\rm A}$}

\newcommand{\mh}{$M_{\rm mol}$} \newcommand{\mhms}{$M_{\rm
    mol}$/$M_\ast$}

\received{}  \revised{}  \accepted{} \submitjournal{ApJ}

\shorttitle{Estimating molecular gas mass of low-z galaxies}
\shortauthors{Y. Gao et al.}

\begin{document}

\title{Estimating the molecular gas mass of low-redshift galaxies from
  a combination of mid-infrared luminosity and optical properties}

\correspondingauthor{Ting Xiao}  \email{xiaoting@zju.edu.cn,
  gaoyang@shao.ac.cn}

\author{Yang Gao} \affiliation{Shanghai Astronomical  Observatory,
  Chinese Academy of Sciences, Shanghai 200030, China}
\affiliation{University of Chinese Academy of Sciences, No.19A Yuquan
  Road, Beijing 100049, China}

\author{Ting Xiao} \affiliation{Department of Physics, Zhejiang
  University, Hangzhou, Zhejiang 310027, China} \affiliation{Shanghai
  Astronomical  Observatory, Chinese Academy of Sciences, Shanghai
  200030, China}

\author{Cheng Li} \affiliation{Department of Astronomy, Tsinghua
  University, Beijing 100084, China}

\author{Xue-Jian Jiang} \affiliation{Purple Mountain Observatory \&
  Key Lab. of Radio Astronomy, Chinese Academy of Sciences, Nanjing
  210034, China}

\author{Qing-Hua Tan} \affiliation{Purple Mountain Observatory \& Key
  Lab. of Radio Astronomy, Chinese Academy of Sciences, Nanjing
  210034, China}

\author{Yu Gao} \affiliation{Purple Mountain Observatory \& Key
  Lab. of Radio Astronomy, Chinese Academy of Sciences, Nanjing
  210034, China}

\author{Christine D. Wilson} \affiliation{Department of Physics and
  Astronomy, McMaster University, Hamilton, ON L8S 4M1, Canada}

\author{Martin Bureau} \affiliation{Sub-department of Astrophysics,
  University of Oxford, Denys Wilkinson Building, Keble Road, Oxford
  OX1 3RH, UK} \affiliation{Yonsei Frontier Lab and Department of
  Astronomy, Yonsei University, 50 Yonsei-ro, Seodaemun-gu, Seoul
  03722, Republic of Korea}

\author{Am\'elie Saintonge} \affiliation{Department of Physics \&
  Astronomy, University College London, Gower Street, London WC1E 6BT,
  UK}

\author{Jos\'e R.~S\'anchez-Gallego} \affiliation{Department of
  Astronomy, Box 351580, University of Washington, Seattle, WA 98195,
  USA}

\author{Toby Brown} \affiliation{Department of Physics and Astronomy,
  McMaster University, Hamilton, ON L8S 4M1, Canada}

\author{Christopher J. R. Clark} \affiliation{Space Telescope Science
  Institute, 3700 San Martin Drive, Baltimore, Maryland, 21218, USA}

\author{Ho Seong Hwang} \affiliation{Korea Astronomy and Space Science
  Institute, 776 Daedeokdae-ro, Yuseong-gu, Daejeon 34055, Republic of
  Korea}

\author{Isabella Lamperti} \affiliation{Department of Physics \&
  Astronomy, University College London, Gower Street, London WC1E 6BT,
  UK}

\author{Lin Lin} \affiliation{Shanghai Astronomical  Observatory,
  Chinese Academy of Sciences, Shanghai 200030, China}

\author{Lijie Liu} \affiliation{Sub-department of Astrophysics,
  University of Oxford, Denys Wilkinson Building, Keble Road, Oxford
  OX1 3RH, UK}

\author{Dengrong Lu} \affiliation{Purple Mountain Observatory \& Key
  Lab. of Radio Astronomy, Chinese Academy of Sciences, Nanjing
  210034, China}

\author{Hsi-An Pan} \affiliation{Institute of Astronomy and
  Astrophysics, Academia Sinica, AS/NTU Astronomy-Mathematics
  Building, No.1, Sec. 4, Roosevelt Rd, Taipei 10617, Taiwan}

\author{Jixian Sun} \affiliation{Purple Mountain Observatory \& Key
  Lab. of Radio Astronomy, Chinese Academy of Sciences, Nanjing
  210034, China}

\author{Thomas G. Williams} \affiliation{School of Physics \&
  Astronomy, Cardiff University, Queens Buildings, The Parade,
  Cardiff, CF24 3AA, UK}

\begin{abstract}
We present CO(J=1-0) and/or CO(J=2-1) spectroscopy for 31 galaxies
selected from the ongoing Mapping Nearby Galaxies at Apache Point
Observatory (MaNGA) survey, obtained with multiple telescopes.  This
sample is combined with CO observations from the literature to study
the correlation of the CO luminosities ($L_{\rm CO(1-0)}$) with the
mid-infrared luminosities at 12 ($L_{\mbox{12\micron}}$) and  22
\micron\ ($L_{\mbox{22\micron}}$), as well as the dependence of the
residuals on a variety of  galaxy properties. The correlation with
$L_{\mbox{12\micron}}$ is tighter and more linear, but galaxies with
relatively low stellar masses ($M_\ast\lesssim10^{10}$ \msolar) and
blue colors  ($g-r\lesssim0.5$ and/or \nuvr\ $\lesssim3$) fall
significantly below the mean $L_{\rm CO(1-0)}$--$L_{\mbox{12\micron}}$
relation. We propose a new estimator of the CO(1-0) luminosity (and
thus the total molecular gas mass \mh) that is a linear combination of
three parameters:	 $L_{\mbox{12\micron}}$, $M_\ast$ and
$g-r$. We show that, with a scatter of only 0.18 dex in log $(L_{\rm
  CO(1-0)})$,  this estimator provides unbiased estimates   for
galaxies of different properties and types. An immediate application
of this estimator to  a compiled sample of galaxies with only
CO(J=2-1) observations yields a distribution of the CO(J=2-1) to
CO(J=1-0) luminosity ratios ($R21$)  that agrees well with the
distribution of real observations, in terms of both the median and the
shape. Application of our estimator to the current MaNGA
  sample reveals a gas-poor population of galaxies that are predominantly
  early-type and show no correlation between molecular gas-to-stellar
  mass ratio and star formation rate, in contrast to gas-rich galaxies.
  We also provide alternative estimators with similar scatters,
based on $r$ and/or $z$ band luminosities instead of $M_\ast$. These
estimators serve as  cheap and convenient \mh\ proxies to be
potentially applied to large samples of galaxies, thus allowing
statistical studies of gas-related processes of galaxies.
\end{abstract} 



\section{Introduction}
\label{sec:introduction}

In current galaxy formation models, galaxies form at the centers  of
dark matter haloes, where gas is able to cool, condense and form stars
\citep[e.g.][]{White1978}.  It is thus crucial to understand  the
physical processes that regulate gas accretion and cycling in/around
galaxies before one can have a complete picture of  galaxy formation
and evolution. Despite of a rich history of studies, however,
  our understanding of the cold gas content of galaxies has been
  rather limited due to the lack of large surveys at radio/mm/sub-mm
  wavelengths.  Large surveys aiming to detect H{\sc i} in nearby
  galaxies have become available only in recent years, such as the
  H{\sc i} Parkes All-Sky Survey  \citep[HIPASS;][]{Zwaan2005} and the
  Arecibo Legacy Fast ALFA (ALFALFA) survey \citep{Giovanelli2005}.
  For molecular gas content, there have also been recent efforts of
  establishing uniform samples of CO detections for nearby galaxies,
  such as the CO Legacy Data base for the GASS \citep[COLD
    GASS;][]{Saintonge2011} and the extended COLD GASS \citep[xCOLD
    GASS;][]{Saintonge2017} surveys.  Unfortunately, when compared to
  optical surveys, these surveys are still relatively shallow and
  small, limited to low redshifts (mostly $z<0.2$) and with poor
  spatial resolution.

In order for statistical studies of both the stellar and
  gaseous content of galaxies, there have been many attempts to
  estimate the cold gas content (both H{\sc i} and H$_2$ mass) for
  large samples of optically-detected galaxies, using galaxy
  properties that can be more easily obtained.  The current H{\sc i}
surveys, together with compiled catalogs of H{\sc i} detections from
the literature \citep[e.g. HyperLeda;][]{Paturel2003}, have revealed
that the H{\sc i} gas-to-stellar mass ratio  ($M_{\rm HI}/M_\ast$)
correlates with a variety of galaxy properties, including specific
star formation rate (sSFR) and related parameters such as optical,
optical--near-infrared (NIR) and near-ultraviolet (NUV)--optical
colors (e.g. \citealp{Kannappan2004,Zhang2009,Catinella2010}), as well
as structural parameters such as  stellar light or mass surface
density (e.g. \citealp{Zhang2009,Li2012}).  Such scaling relations
have motivated many attempts to calibrate colors, H$\alpha$
luminosity, or a combination of multiple parameters  as proxies for
$M_{\rm HI}/M_\ast$, providing  {\em estimated} H{\sc i} masses for
large samples of galaxies,  thus allowing statistical studies of
gas-related processes \citep[e.g.][]{Kannappan2004, Tremonti2004,
  Erb2006, Zhang2009, Catinella2010, Li2012, Brinchmann2013,
  Kannappan2013, Zhang2013, Eckert2015, Rafieferantsoa2018,
  Zu2018}. Such estimators typically have a scatter of $\sim
0.25-0.4$ dex in log$(M_{\rm HI}/M_\ast)$.

As pointed out in \citet[][see their Section 3.2]{Zhang2009}, such
H{\sc i} gas mass estimators can be understood from the
Kennicutt-Schmidt (KS) star formation relation
\citep{Schmidt1959,Kennicutt1998}, that relates the star formation
rate per unit area ($\Sigma_{\rm SFR}$) to the surface mass density of
cold gas ($\Sigma_{\rm gas}$) in a galactic disc.  Because star
formation is expected to occur in cold giant molecular clouds
\citep{Solomon1987,McKee2007,Bolatto2008}, one might expect the CO
(and \hh) emission to present tighter correlations with SFR-related
properties than the H{\sc i} emission.  Indeed, the KS law with
molecular gas surface densities measured from CO emission is found to
be more linear (with a slope closer to unity) than that from H{\sc i}
emission \citep[e.g.][]{Bigiel2008, Leroy2008}.  Meanwhile,
\citet{Gao2004} find a tight linear relation between the integrated
SFRs and dense molecular gas masses  (derived from HCN emission) of
normal and starburst galaxies.  Combined with some CO observations at
high redshifts \citep[e.g.][]{Carilli2013,Riechers2019}, previous
studies have also shown that the molecular gas content of galaxies is
well correlated with  the cosmic star formation rate density
\citep[e.g.][]{Kruijssen2014,Saintonge2017,Tacconi2018}. In addition,
the ratio of \hh\ mass (inferred from CO emission observations) to
stellar mass ($M_{\rm H_2}/M_\ast$) is found to correlate with sSFR
and \nuvr\ of nearby galaxies, as nicely shown by the COLD GASS
\citep{Saintonge2011} and xCOLD GASS \citep{Saintonge2017}
surveys. However, in the same surveys, $M_{\rm H_2}/M_\ast$ shows only
a mild dependence on stellar mass at $M_\ast<10^{10.5}$ \msolar\ and
on stellar surface mass density at $\mu_\ast\lesssim10^{8.7}$
\msolar\ kpc$^{-2}$, before it drops sharply at higher masses and/or
surface densities. This behavior is in contrast to $M_{\rm
  HI}/M_\ast$, that decreases quasi-linearly with increasing $M_\ast$
or  $\mu_\ast$ down to $M_{\rm HI}/M_\ast\sim0.01$ at log$(\mu_\ast $/
(\msolar\ kpc$^{-2}$)) $\sim 9.6$ \citep[e.g.][see their
  Fig. 2]{Zhang2009}.

Apparently, the molecular gas content of galaxies doesn't
  scale with their optical properties in a simple way.  Many studies
have attempted to link the  molecular gas content of galaxies with
their infrared luminosities.  For instance, far-infrared (FIR) or
sub-mm continuum  observations are commonly used to derive total dust
masses, from which total gas masses are inferred with the
(metallicity-dependent) gas-to-dust mass ratio
(e.g. \citealp{Israel1997, Leroy2011, Magdis2011, Eales2012,
  Sandstrom2013, Scoville2014, Groves2015}).  This gas-to-dust ratio
is relatively constant,  considering the extremely large CO-to-H2
conversion factor in extremely metal-poor galaxies \citep{Shi2016}.  A
new survey, JINGLE (JCMT dust and gas In Nearby Galaxies Legacy
Exploration), is obtaining both  integrated CO line spectroscopy and
850 \micron\ continuum fluxes for nearby galaxies using the 15-m James
Clark Maxwell Telescope (JCMT). It will study the scaling relations of
cold gas and dust masses with global galaxy properties such as stellar
mass, SFR and gas-phase metallicity \citep{Saintonge2018}.

Furthermore, thanks to the \textit{Wide-field Infrared Survey
  Explorer}  \citep[\textit{WISE};][]{Wright2010}, mid-infrared (MIR)
luminosities have recently been found to strongly correlate with CO
emission in both nearby star-forming late-type and (generally
non-star-forming) early-type galaxies \citep{Kokusho2017,Kokusho2019}.
Some authors found \textit{WISE} 4.6-12 \micron\ color to
  strongly correlate with star formation activity \citep{Donoso2012}
  and molecular gas mass fraction \citep{Yesuf2017}.  By jointly
analyzing the \textit{WISE} data and the CO observations from COLD
GASS, Analysis of the interstellar Medium of Isolated GAlaxies
\citep[AMIGA;][]{Lisenfeld2011} and their own sample observed with
Sub-millimeter Telescope (SMT), \citet{Jiang2015} found both  the
CO(J=1-0) and CO(2-1) luminosities ($L_{\rm CO(1-0)}$ and $L_{\rm
  CO(2-1)}$) to tightly correlate with the W3 (12 \micron) luminosity
($L_{\mbox{12\micron}}$), the relation being well described by a power
law with a slope close to unity for CO(J=2-1) and $\sim 0.9$ for
CO(J=1-0).  These relations are anticipated, as the authors pointed
out, considering the previous finding that the majority ($\sim 80$\%)
of the  12 \micron\ emission from star-forming galaxies in
\textit{WISE} is produced by stars younger than $\sim 0.6$ Gyr
\citep{Donoso2012}. Therefore, the 12 \micron\ luminosity, that is
available from the \textit{WISE} all-sky catalogue,  can be adopted as
a cheap and convenient estimator of CO luminosity for galaxies. In
fact, this single-parameter estimator as proposed in \citet{Jiang2015}
has been adopted to estimate the observing times for target selection
for the JINGLE project \citep{Saintonge2018}.

The tight correlation between CO luminosities and the 12
  \micron\ luminosity as found by \citet{Jiang2015} can in principle
  be applied to large samples of galaxies, thus enabling star
  formation and gas-related processes to be studied
  statistically. However, before performing such analyses, one might
  wonder whether and how this molecular gas mass estimator can be
  further improved.  In this work we present CO (J=1-0) and/or CO
(J=2-1) observations of 31 galaxies selected from the ongoing Mapping
Nearby Galaxies at Apache Point Observatory (MaNGA) survey
\citep{Bundy2015}, obtained using the Purple Mountain Observatory
(PMO) 13.7-m millimeter telescope located in Delingha, China,  the
JCMT and the 10.4-m Caltech Submillimeter Observatory (CSO) in
Hawaii. We extend the work of \citet{Jiang2015} by combining
our sample with public data from xCOLD GASS, AMIGA and the Herschel
Reference Sample \citep[HRS;][]{Boselli2014}, and studying
the residuals about the CO versus MIR luminosity relation as a
function of various galaxy properties. 

The purpose of our work is multifold. First, we attempt to extend the
$L_{\rm CO}$-$L_{\mbox{12\micron}}$ relation by including one or  more
parameters, in the hope of finding a better and unbiased estimator of
$L_{\rm CO}$ (thus the total molecular gas mass \mh) for future
applications.  As we will show, once combined with one
  additional property from optical observations, the 12
  \micron\ luminosity can more accurately predict the CO (1-0)
  luminosity of galaxies with a scatter \textless\ 0.2dex and no
  obvious biases.  Second, the PMO 13.7-m telescope has so far been
mainly dedicated to Galactic observations, and it is important to
determine under what conditions this telescope may also be useful for
extra-galactic observations.  Third, the JCMT-based observations of
this work were proposed initially as a pilot for the JINGLE
project. All observations presented in this work should thus be
complementary to the JINGLE CO observations. Finally, our galaxies are
selected from MaNGA, and the integral field spectroscopy will  allow
us to link the global measurements of cold gas with spatially-resolved
stellar and ionized gas properties. This will be the topic of our next
work. In the current paper we will present an application of
  our estimator to the current sample of MaNGA, and examine the
  correlation of molecular gas-to-stellar mass ratio
  (${\mbox{M}}_{\mbox{mol}}/{\mbox{M}}_\ast$) with star formation rate
  (SFR) for different classes of galaxies.

In the next section we describe our sample selection,
  observations and data reduction. In Section~\ref{sec:co_mir} we examine
  the correlations  of CO luminosities with mid-infrared
  luminosities. We present a new estimator of the CO(J=1-0)
  luminosity, as well as a simple application of the estimator to
  derive the CO(2-1)-to-CO(1-0) line ratio for  a sample of local
  galaxies. In Section~\ref{sec:mol_gas} we calculate the molecular gas
  mass from the observed CO spectra of our galaxies, and apply our CO
  luminosity estimator to the MaNGA sample.  Finally, we summarize
and discuss our work in Section~\ref{sec:summary}.  Throughout the
paper, distance-dependent quantities are calculated  by assuming a
standard flat ${\Lambda}$CDM cosmology with a matter density parameter
$\Omega_{\rm m}=0.275$ , a dark energy density parameter
$\Omega_\Lambda=0.725$ and a Hubble constant of $h$ = 0.7 following
\citet{Komatsu2011}.

\begin{figure*}
\begin{center}
\includegraphics[height=0.4\textwidth]{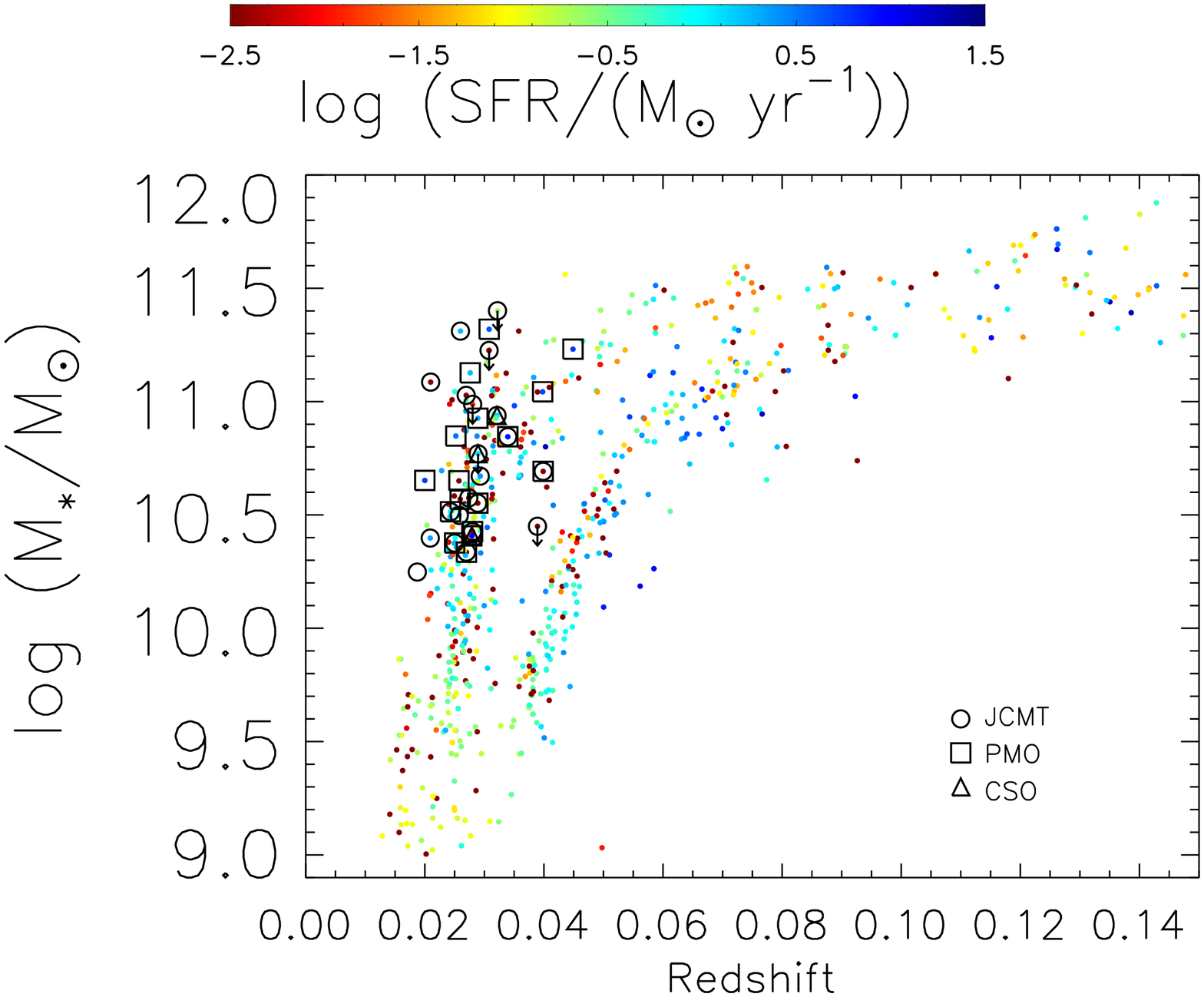}
\includegraphics[height=0.4\textwidth]{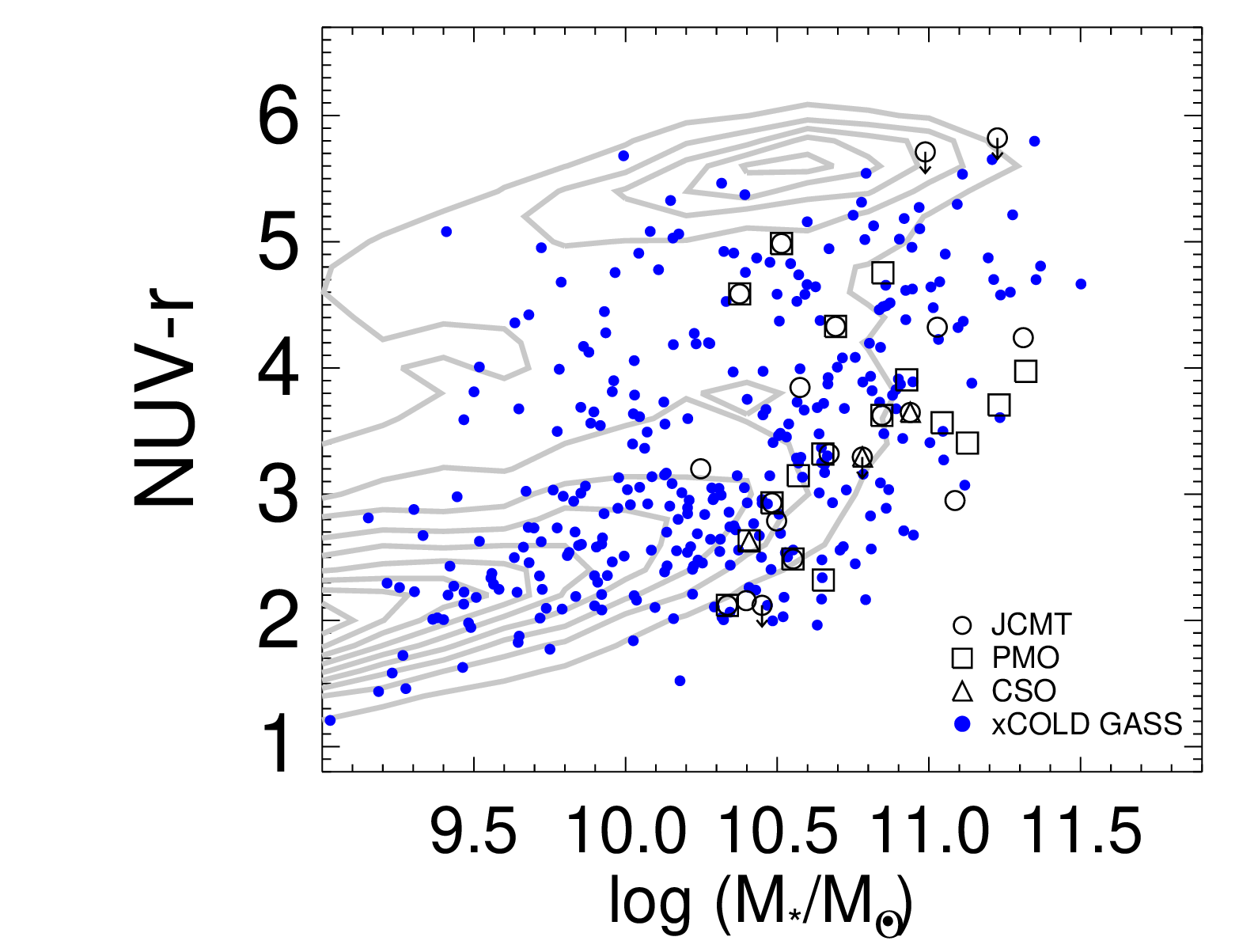}
\caption{Distribution of our target galaxies on the  plane of
    stellar mass versus redshift (left panel) and the plane of {\nuvr}
    color versus stellar mass (right panel).  In both panels, the open
    circles, squares and triangles represent our targets that are
    observed with the JCMT, PMO and CSO respectively.  Symbols with a
    downward arrow represent CO upper limits from our JCMT
    observations.  Small filled circles in the left panel show the
    MaNGA MPL-3 sample for comparison, color-coded by star formation
    rate taken  from \citet{Salim2016}. Blue dots in the right panel
    show CO detections from the xCOLD GASS survey, while the grey
    contours show the distribution of a volume-limited galaxy sample
    selected from the SDSS/DR7 main galaxy sample.  }
\label{fig:sele_sample}
\end{center}
\end{figure*}

\section{Sample and Observations}
\label{sec:data}
\subsection{Target selection}

Our parent sample is selected from the MaNGA Product Launch 3 (MPL-3),
the latest MaNGA sample available when this work was initiated.  The
MPL-3 included $720$ galaxies with redshifts $z<0.15$ and  stellar
masses above $\sim 10^9$ \msolar. It is a random subset  of the 10,000
galaxies planned for the full MaNGA survey \citep{Bundy2015}.   The
MaNGA survey sample was selected from the Sloan Digital Sky Survey
(SDSS) spectroscopic sample, mainly to have a flat distribution of
$i$-band absolute magnitudes and for the assigned integral-field unit
bundles to reach 1.5 or 2.5 effective radii ($R_{\rm e}$). The MaNGA
sample design and optimisation is described in detail in
\citet{Wake2017}.  The left panel of
Figure~\ref{fig:sele_sample} shows all the galaxies from the MaNGA
MPL-3 on the plane of $\log$(\mstar) versus redshift,
color-coded by the star formation rate (SFR) taken from
  \citet{Salim2016}.  The galaxies are distributed on two separate
loci, corresponding to the two radius limits adopted in the MaNGA
sample selection, i.e. 1.5 and 2.5 $R_{\rm e}$.  Considering both the
limited capability of our telescopes and the limited observing time,
we restrict ourselves to  low-redshift massive galaxies with redshifts
$z<0.05$ and stellar masses \mstar\ $>10^{10}$ \msolar.  This
restriction gives rise to a parent sample of 281 galaxies.

We have utilized three telescopes, the PMO 13.7-m telescope, the JCMT
and CSO in Hawaii to obtain CO spectra for our target galaxies.  We
select targets from the same parent sample as described above, but for
the three telescopes independently considering the following aspects.
Firstly, the three telescopes have quite different sensitivities. JCMT
is the most sensitive for a given observing time, while PMO is much
less sensitive and so can only observe the brightest targets.
Secondly, the PMO telescope can observe only the CO(1-0) line, while
JCMT and CSO observe the CO(2-1) line. It is hard to construct a
homogeneous sample of galaxies  by obtaining CO observations for
different galaxies from different telescopes. Therefore, we decided to
first select a parent sample from the MaNGA survey, and then perform
CO observations for three sub-samples of galaxies, each using one of
the three telescopes independently. In this way, we expect to have
some targets that are observed by more than one telescope, and these
observations will be helpful both to cross-check the flux calibration
of the different telescopes and to probe the ratio between the CO(2-1)
and CO(1-0) lines. 

For the PMO 13.7-m telescope, we select 17 galaxies that are
brightest in the 12 \micron, with the \textit{WISE} W3 flux
$f_{\mbox{12\micron}}$\textgreater\ 28 mJy. We detected CO(1-0) emission
in all the 17 galaxies with a signal to noise ratio $S/N \geq$ 3.
For the JCMT, we randomly selected a subset of the parent sample,
but requiring that the total observing time per target required to
reach $S/N = 5$ must not exceed 5 hours, assuming Band 4 weather conditions.
For this purpose we have estimated the observing time for each galaxy
in MaNGA MPL-3 using the $L_\mathrm{CO}$-$L_{\mbox{12\micron}}$ relation
from \citet{Jiang2015}, and randomly select 21 out of 49 galaxies that
meet the requirements. We detected CO(2-1) emission with $S/N \geq$ 3
in 16 galaxies, of which 7 are also observed with the PMO 13.7-m telescope.
Finally, the CSO was used to obtain additional CO(2-1) observations
for a small number of galaxies that are randomly selected from
the parent sample without considering the observations at the other
two telescopes.  Due to the limited allocated time, only
three galaxies are observed with this telescope, of which one is also
observed with PMO, and two with JCMT. These observations are described
in more detail in Section~\ref{sec:co_obs}.

In summary, we have observed a sample of 31 galaxies using the three
telescopes, of which 27 yielded a reliable detection.  These 31 galaxies
are highlighted in Figure~\ref{fig:sele_sample}  as black symbols, and their 
general properties  including redshift, stellar mass, UV-to-optical
color and infrared luminosity are listed in Table~\ref{tab:tbl1}.
In the right panel of Figure~\ref{fig:sele_sample},
we show the 31 galaxies in our sample on the plane of \nuvr\ versus
stellar mass. For comparison, we also show the xCOLD GASS sample
galaxies \citep{Saintonge2017} as blue dots which include 532 galaxies
with CO(1-0) measurements from the IRAM (Institut de Radioastronomie Millim\'etrique)
30-m telescope, as well as a volume-limited galaxy sample selected
from the SDSS main galaxy sample as grey contours which 
consists of 33,208 galaxies with redshifts $0.01<z<0.03$ and stellar masses
$M_\ast>10^{9}$ \msolar.
Stellar masses, \nuvr\ colors and redshifts for all galaxies in Figure~\ref{fig:sele_sample}
are taken from the NASA Sloan Atlas (NSA), a catalogue of images and parameters
of more than 640,000 galaxies with $z<0.15$ from SDSS, as described in
detail in \citet{Blanton2011}.

As can be seen from Figure~\ref{fig:sele_sample}, our sample is limited
to relatively massive galaxies with $M_\ast\ga 2\times 10^{10}$ \msolar.
Our sample is also biased to blue colors when compared to the general
population of galaxies from SDSS,  but it appears to cover the \nuvr\
space in a similar way to the galaxies detected in the xCOLD GASS survey.

\begin{deluxetable*}{rlrrrrr}
\tablecaption{ List of targets and their general properties.}
\label{tab:tbl1}
\tablewidth{0pt}
\tablehead{
\colhead{Target No.} & \colhead{SDSS ID}  & \colhead{$z$} & 
\colhead{$\log({\rm M}_\ast/{\rm M}_\odot)$}  & \colhead{NUV$-r$} & 
\colhead{$\log({\rm L}_{12\micron}/{\rm L}_\odot)$} & \colhead{$\log({\rm L}_{22\micron}/{\rm L}_\odot)$} \\
\colhead{} & \colhead{} & \colhead{} & \colhead{} & \colhead{(mag)} & \colhead{} & \colhead{}
}
\decimalcolnumbers
\startdata
1	&	J093637.19+482827.9	&	0.026	&	10.57	&	3.15	$\pm$	0.06	&	9.80	$\pm$	0.02	&	10.74	$\pm$	0.03	\\
2	&	J093106.75+490447.1	&	0.034	&	10.85	&	3.62	$\pm$	0.07	&	9.82	$\pm$	0.02	&	10.77	$\pm$	0.03	\\
3	&	J091554.70+441951.0	&	0.040	&	11.04	&	3.57	$\pm$	0.05	&	10.11	$\pm$	0.02	&	11.00	$\pm$	0.03	\\
4	&	J032057.90-002155.9	&	0.021	&	11.09	&	2.95	$\pm$	0.06	&	9.40	$\pm$	0.02	&	9.88	$\pm$	0.03	\\
5	&	J110158.99+451340.9	&	0.020	&	10.65	&	2.32	$\pm$	0.05	&	9.46	$\pm$	0.02	&	10.22	$\pm$	0.03	\\
6	&	J091500.75+420127.8	&	0.028	&	10.41	&	2.63	$\pm$	0.05	&	9.59	$\pm$	0.02	&	10.41	$\pm$	0.03	\\
7	&	J032043.20-010633.0	&	0.021	&	10.40	&	2.16	$\pm$	0.05	&	9.10	$\pm$	0.03	&	9.67	$\pm$	0.03	\\
8	&	J110637.35+460219.6	&	0.025	&	10.38	&	4.59	$\pm$	0.06	&	9.37	$\pm$	0.03	&	10.33	$\pm$	0.03	\\
9	&	J141225.99+454129.9	&	0.027	&	10.34	&	2.12	$\pm$	0.06	&	9.34	$\pm$	0.03	&	10.15	$\pm$	0.03	\\
10	&	J152950.65+423744.1	&	0.019	&	10.25	&	3.20	$\pm$	0.05	&	8.64	$\pm$	0.03	&	9.24	$\pm$	0.03	\\
11	&	J073749.42+462351.5	&	0.032	&	10.94	&	3.65	$\pm$	0.05	&	9.39	$\pm$	0.03	&	9.86	$\pm$	0.03	\\
12	&	J074442.28+422129.3	&	0.039	&	10.45	&	2.12	$\pm$	0.05	&	9.45	$\pm$	0.03	&	10.09	$\pm$	0.03	\\
13	&	J211557.49+093237.9	&	0.029	&	10.67	&	3.32	$\pm$	0.08	&	9.44	$\pm$	0.02	&	9.99	$\pm$	0.03	\\
14	&	J090015.61+401748.3	&	0.029	&	10.93	&	3.91	$\pm$	0.07	&	9.34	$\pm$	0.03	&	10.01	$\pm$	0.03	\\
15	&	J152625.50+422114.0	&	0.028	&	10.48	&	2.93	$\pm$	0.05	&	9.38	$\pm$	0.03	&	9.89	$\pm$	0.03	\\
16	&	J031345.21-001429.2	&	0.026	&	11.31	&	4.24	$\pm$	0.05	&	9.22	$\pm$	0.03	&	9.64	$\pm$	0.03	\\
17	&	J221134.29+114744.9	&	0.027	&	11.03	&	4.32	$\pm$	0.05	&	9.17	$\pm$	0.03	&	9.71	$\pm$	0.03	\\
18	&	J220943.19+133802.9	&	0.027	&	10.58	&	3.85    \tablenotemark{a}    &	9.46	$\pm$	0.02	&	10.01	$\pm$	0.03	\\
19	&	J171100.29+565600.9	&	0.029	&	10.55	&	2.49	$\pm$	0.05	&	9.38	$\pm$	0.03	&	10.01	$\pm$	0.03	\\
20	&	J093813.89+482317.9	&	0.026	&	10.50	&	2.79	$\pm$	0.06	&	9.09	$\pm$	0.03	&	9.58	$\pm$	0.03	\\
21	&	J072333.23+412605.6	&	0.028	&	11.13	&	3.41	$\pm$	0.06	&	9.32	$\pm$	0.03	&	9.80	$\pm$	0.03	\\
22	&	J110704.16+454919.6	&	0.025	&	10.85	&	4.75	$\pm$	0.06	&	9.12	$\pm$	0.03	&	9.73	$\pm$	0.03	\\
23	&	J092138.71+434334.2	&	0.040	&	10.69	&	4.33	$\pm$	0.06	&	9.74	$\pm$	0.03	&	10.33	$\pm$	0.03	\\
24	&	J121336.85+462938.2	&	0.026	&	10.65	&	3.32	$\pm$	0.05	&	9.36	$\pm$	0.03	&	9.81	$\pm$	0.03	\\
25	&	J074637.70+444725.8	&	0.031	&	11.32	&	3.97	$\pm$	0.05	&	9.51	$\pm$	0.02	&	9.94	$\pm$	0.05	\\
26	&	J134145.21+270016.9	&	0.029	&	10.78	&	3.29	$\pm$	0.07	&	9.05	$\pm$	0.03	&	9.40	$\pm$	0.04	\\
27	&	J103731.86+433913.7	&	0.024	&	10.51	&	4.99	$\pm$	0.06	&	9.12	$\pm$	0.03	&	9.54	$\pm$	0.04	\\
28	&	J083445.04+524256.4	&	0.045	&	11.23	&	3.71	$\pm$	0.05	&	9.69	$\pm$	0.03	&	10.04	$\pm$	0.04	\\
29	&	J103038.52+440045.7	&	0.028	&	10.99	&	5.71	$\pm$	0.15	&	8.36	$\pm$	0.05	&	8.40	$\pm$	0.11	\\
30	&	J171523.26+572558.3	&	0.032	&	11.40	&	12.62	\tablenotemark{a}&	8.13	$\pm$	0.06	&	$\dots$			\\    
31	&	J110310.99+414219.0	&	0.031	&	11.23	&	5.82	$\pm$	0.13	&	8.85	$\pm$	0.06	&	$\dots$			\\
\enddata
\tablecomments{From left to right, the columns are: (1) serial number unique to
          each target and kept the same in Table~\ref{tab:tbl2}; (2) SDSS
          name formed by the R.A. and Dec. of the target; (3) optical redshift from SDSS;
(4) \& (5) stellar mass (a detailed discussion in Section~\ref{sec:alter}) and \nuvr\ color (and its uncertainty) from NSA; (6) \& (7) mid-infrared luminosities
          at 12 and 22 \micron\ (and their uncertainties) as measured by ourselves from the \textit{WISE}
          images. There are two non-detections at 22\micron.}
\tablenotetext{a}{The NUV$-r$ of these two galaxies is unreliable.}
\end{deluxetable*}

 \subsection{CO Observations} \label{sec:co_obs}

Our observations of the CO(1-0) emission line at the PMO 13.7-m
telescope were carried out over two periods in 2015,  one from May 5th
to June 1st, the other from November 6th to December 25th.  These
observations were taken with a nine-beam superconducting spectroscopic
array receiver \citep{Shan2012} with a main beam efficiency $\eta_{\rm
  mb} = 0.513$ and a half--power beam width (HPBW) $\theta_{\rm HPBW}
\sim 52''$. We have examined the optical size of our galaxies, as quantified by R$_{90}$ from the NSA \citep{Blanton2011}, the radius enclosing 90\% of the total light in r-band, and we found more than 75\% of our galaxies have a R$_{90}$ that is smaller than half of the HPBM of the PMO 13.7m telescope.  In practice, the observations were made in
ON-and-OFF mode.  For each target galaxy, two of the nine beams were
used, one covering the target and one pointing to an off-target area,
thus simultaneously obtaining two short scans (each of 1 minute) of
both the target and the background, that are then switched between the
two beams.  For each galaxy, we then combined a selected set of
reliable scans to obtain the final integrated spectrum.  For this, we
visually examined all the scans, and discarded the spectra  of scans
with strongly distorted baselines,  extremely large noise and/or any
anomalous feature (mainly due to bad weather or high system
temperature). We find the fraction of usable scans is $\sim$60\% when
the system temperature $T_{\rm sys}\lesssim 200$K, and decreases
sharply at higher temperatures. The effective on-source time is $\sim$
75 hours for the observations in the winter period, but only 95
minutes for those in May, although the  actual allocated time was much
longer in the latter period. 

The observations of CO(2-1) at JCMT were taken from
March to November in 2015 (project codes: M15AI28 and M15BI060; PI: Ting
Xiao) with the RxA receiver in weather band 4 and 5. In total 16
hours of on-source time were allocated to the 21 target galaxies.  RxA
is a single receiver dual sideband (DSB) system, covering  the
frequency range 212 to 274 GHz, with $\eta_{\rm mb} = 0.65$  and 
$\theta_{\rm HPBW} \sim 20''$  at 230 GHz.  During these
observations, the opacity $\tau$ at 225 GHz was  less than 0.5. Once fully reduced, 
the observations yielded 16 detections and 5 non-detections for the 21 targeted galaxies.

At CSO, the CO(2-1) observations were obtained for the three target galaxies
on 2015 February 19 and 20, using the heterodyne
receiver with a full--width at half--maximum beam size of $30.\!{''}3 \times 30.\!{''}7$ and a
main beam efficiency of 76$\%$  at 230 GHz. For our observations, the
typical system temperature  ranged from 250 to 340 K, and the opacity
$\tau$ was less than 0.2.

\begin{figure*}
  \begin{center}
    \epsfig{figure=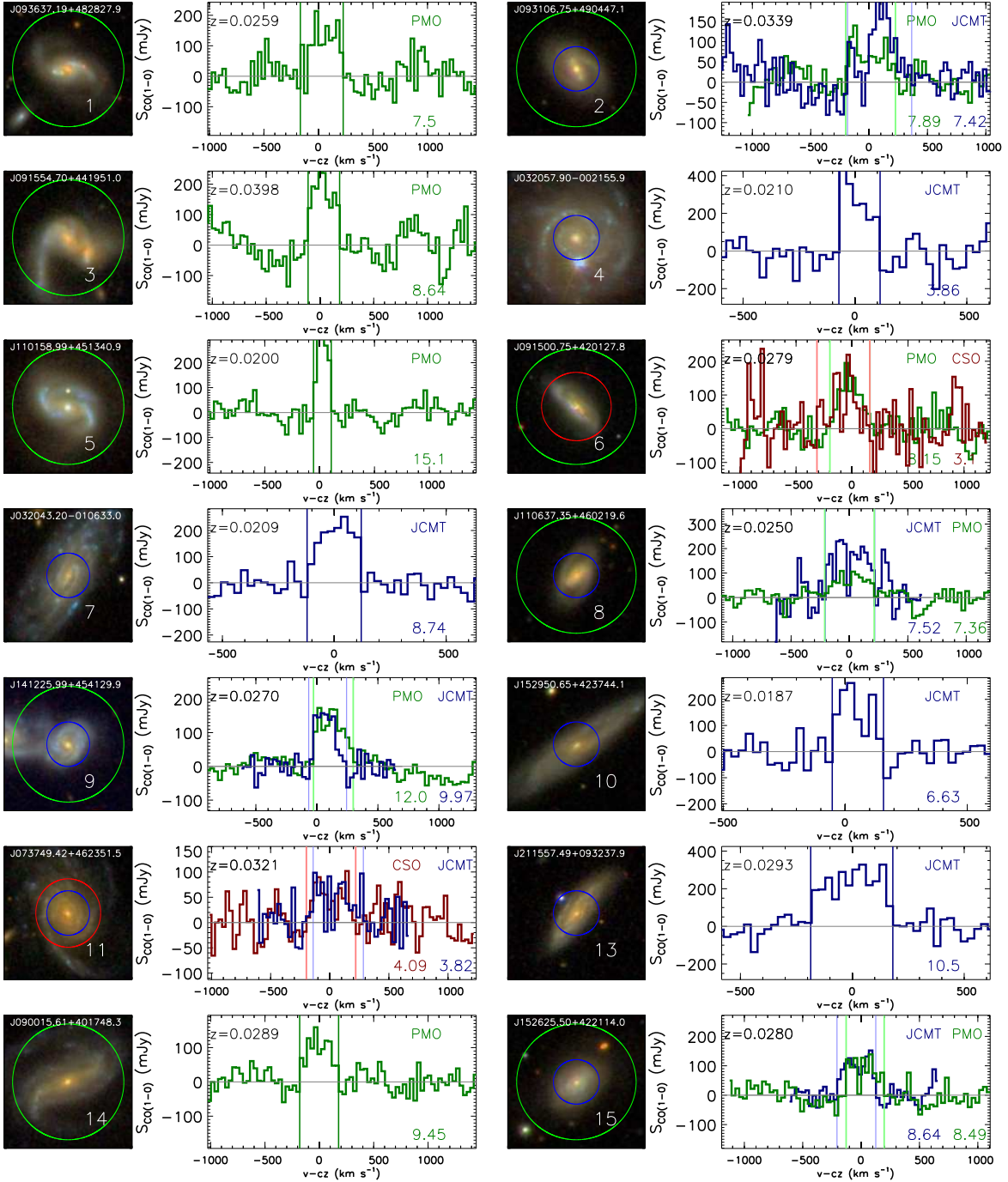,clip=true,width=0.9\textwidth}
  \end{center}
   \caption{SDSS color images and CO spectra of our target galaxies. Each galaxy's 
     SDSS name and target number are overlaid on each image, while the telescope
     used and the $S/N$ are noted in the top-right and bottom-right corner of each spectrum.
     The green, blue and red circles overlaid on the images show the beam size of respectively PMO, JCMT and CSO as relevant. 
     We used PMO to observe CO(1-0) and 
     JCMT and CSO to observe CO(2-1). To directly compare the intensities from different telescopes, all spectra are in
     $S_{\rm CO(1-0)}$ units (assuming $R21$ = 0.7) and binned into channels
     $\sim$ 30 km s$^{-1}$ wide. The velocity scale displayed was obtained by subtracting the systemic velocity (redshift) from NSA catalogue.
     The vertical lines overplotted on the spectra indicate the velocity ranges (FWZI) of the CO emissions.}  
\label{fig:co_spectr}
\end{figure*}

\addtocounter{figure}{-1}
 \begin{figure*} 
  \begin{center}
    \epsfig{figure=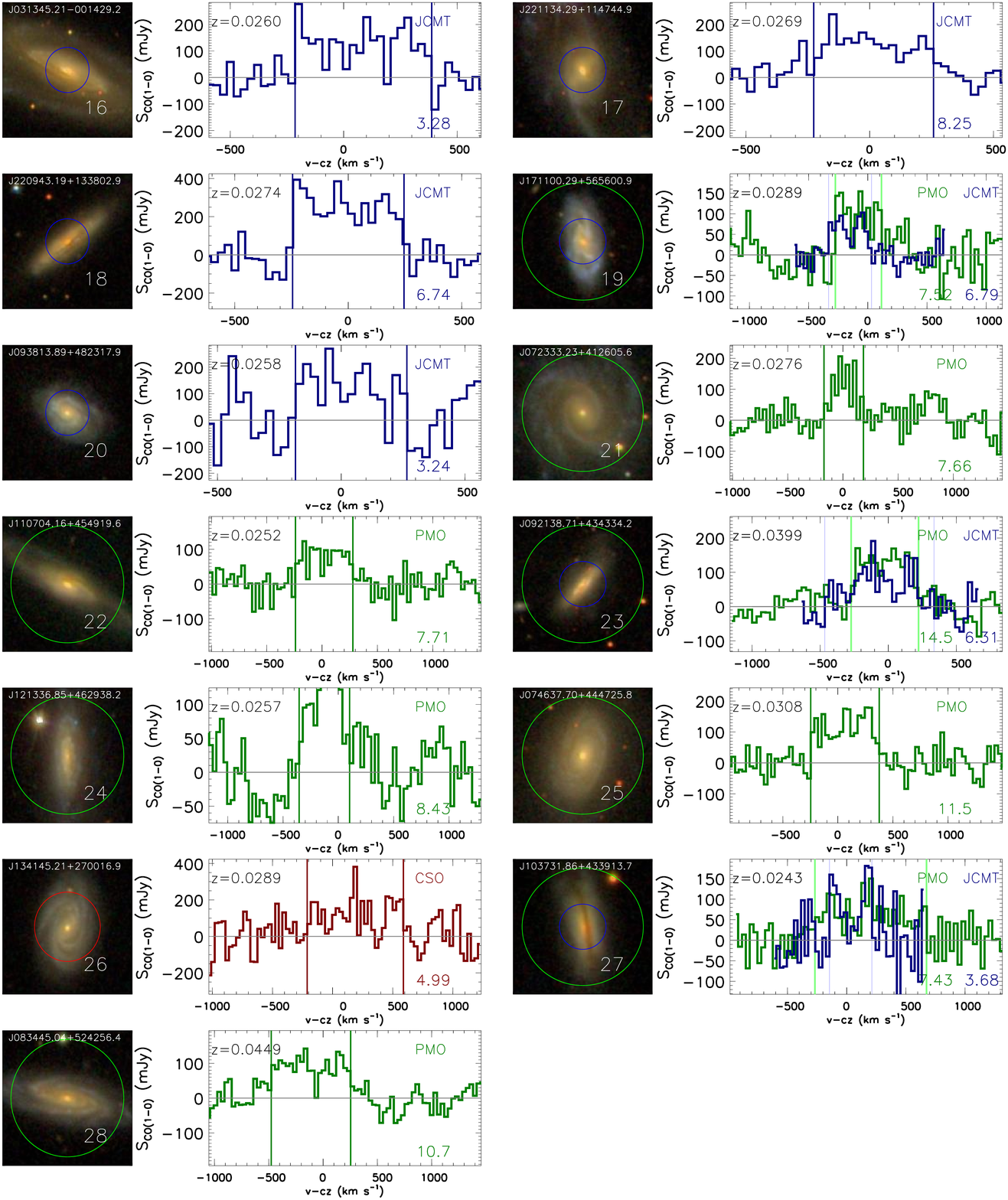,clip=true,width=0.9\textwidth}
  \end{center}
   \caption{(Continued) }
\end{figure*}

\subsection{Data Reduction}

We use the {\tt CLASS} package to reduce the data obtained at PMO and
CSO, part of the {\tt GILDAS} software package \citep{Pety2005}. As
described above, we visually examined all the scans and discarded
unusable scans. In some of the selected scans, there are abnormally
strong "line-like" features (stronger than 5 $\sigma$), where $\sigma$
is the root mean square (rms) noise.  These spikes appear in
  individual original scans with channel width of $\delta v \sim$
  0.16 km s$^{-1}$; they contribute very little to the CO emission line
  measurement, but may affect the determination of the baseline.
  We replace the fluxes of the channels with spike features by
  the average flux of the neighbouring channels, following
  \citet{Tan2011}.  For a given galaxy, we then first obtain a linear
baseline for each scan by fitting the spectrum over the full frequency
range (except the expected range of the CO emission line), and
subtract the fit from  the spectrum. All the baseline-subtracted scans
are then stacked to obtain an average spectrum of the galaxy. During
the stacking, the different scans are weighted by the inverse of their
rms noise.  The {\tt STARLINK} package \citep{Currie2014} is used to
reduce the JCMT data, with the default pipeline. We subtract the
baselines to obtain the final spectra, that are then binned to a
channel width of $\delta v \sim$ 30 km s$^{-1}$.  The intensities
are converted to main beam temperature $T_{\rm mb}$ from antenna
temperature  \TA\ using $T_{\rm mb} = T^{*}_{\rm A}/\eta_{\rm mb}$.

If a CO emission feature in the average spectrum appears significant,
we select its velocity range manually as the full--width at
zero--intensity (FWZI). We then measure the velocity-integrated CO
line intensity, $I_{\rm CO} \equiv \int {T_{\rm mb}}\;dv$,  by
integrating the spectrum over this velocity range that reasonably
covers the line feature.  The uncertainty of the integrated intensity
is estimated using the  standard error formula in \citet{Gao1996}: 
\begin{equation}
  \Delta I_{\rm CO} \equiv T_{\rm rms}\Delta v_{\rm FWZI}\;/\;[f(1-\Delta
    v_{\rm FWZI}/W)]^{1/2} , 
\end{equation}
where $T_{\rm rms}$ is the rms noise computed over the full spectrum
(excluding the emission line),  $f \equiv \Delta v_{\rm FWZI} /\delta
v $ where $\Delta v_{\rm FWZI}$ is the FWZI of the emission feature
and  $\delta v $ the velocity channel width,  and $W$ is the entire
velocity coverage of the spectrum. 

In cases where the CO line is undetected ($S/N<$3) or very weak,  few
of the targets have an H{\sc i} line width, so we compute a value of
${\Delta}I_{\rm CO}$ in the same way as above,  but adopting a fixed
FWZI of 300 \kms\ following \citet{Saintonge2011}. For these
non-detections,  an upper limit to the velocity-integrated intensity
is then given as $3 {\Delta}I_{\rm CO}$. 

Assuming these galaxies are point-like sources,  we convert the
velocity-integrated line intensities in main beam brightness
temperature ($T_{\rm mb}$) scale, i.e. $I_{\rm CO}$ as obtained above,
to $S_{\rm CO}$, the CO line flux density in units of Jy km s$^{-1}$
using a conversion factor of 24.9, 18.4 and 40.2 Jy K$^{-1}$ for
PMO, JCMT and CSO, respectively. 
Figure~\ref{fig:co_spectr} displays the SDSS image (with the
  relevant telescope primary beam overlaid) and the final CO spectrum
  for the 27 detected galaxies in our sample. The spectra are
  plotted in terms of $S_{\rm CO (1-0)}$, that is the flux density
  of the CO (J=1-0) line. For the galaxies observed with JCMT or
  CSO, we have converted the CO (2-1) flux density to the
  CO (1-0) flux density assuming a CO(2-1)/CO(1-0) line ratio $R21$ = 0.7 \citep{Leroy2013}.
  In fact, the median $R21$ of the galaxies in our sample that have
  both CO(1-0) and CO(2-1) detections is also $\sim$0.7. For the galaxies observed
by more than one telescope, we plot the spectra from the different
telescopes with different colors. The repeated observations are
generally in good agreement, in terms of both the intensity and width
of the CO line.

Finally, we calculate the CO line luminosity following
\citet{Bolatto2013}:
\begin{equation} \label{eq:co_lum}
 \Big(\frac{L_{\rm CO}}{{\rm K\;km\;s}^{-1}\;{\rm pc^{2}}} \Big) = 2453~ \Big(\frac{S_{\rm CO} \Delta v}{{\rm Jy\;km\;s}^{-1}}\Big) \Big(\frac{D_{\rm L}}{\rm Mpc}\Big)^{2} (1+z)^{-1},
\end{equation}
where $S_{\rm CO}\Delta v$ is the velocity integrated CO line flux
density, $D_{\rm L}$ is the luminosity  distance to the source, and
$z$ is the source redshift from SDSS.  The CO luminosity and
  corresponding uncertainty for each galaxy in our sample  are listed
  in Table~\ref{tab:tbl2}.  For the galaxies observed by more than
one telescope, we list the results from all the telescopes.  In total,
there are 41 observations including 36 detections for 27 galaxies, and
5 non-detections for 5 galaxies (1 galaxy has both a detection and a
non-detection). Of the 31 galaxies, 10 were repeatedly observed: 7
with both PMO and JCMT, 2 with both JCMT and CSO, and 1 with both PMO
and CSO.  One galaxy with non-detection has a very blue color
(\nuvr\ $=2.12$) and a stellar mass of $M_\ast=10^{10.45}$
\msolar\ (target No. 11). This non-detection should be attributed to
the relatively short integration of the observation due to our limited
observing time.

We have attempted to correct the CO luminosities for the effect of the 
limited apertures of the
telescopes. This aperture effect is negligible
for the CO(1-0) observations obtained with  the PMO 13.7-m telescope, 
that has a rather large beam size, larger than the optical diameter of our
galaxies. In fact, following \citet{Saintonge2012},  we estimated
the ratio between the predicted total CO(1-0) flux and the flux observed within
the beam, finding a median difference of less than 5\% for PMO
galaxies. Therefore, their CO flux can be nearly perfectly recovered with a
single pointing, and we choose to not make a correction for these
observations. For the CO(2-1)  observations obtained with JCMT and/or
CSO, the aperture effect cannot be ignored. We have thus performed
an aperture correction for these observations adopting the method of
\citet{Saintonge2012}, based on the assumption that molecular
gas within galaxies follows the same exponential profile as
their stellar light. The correcting factors range from 1.05 to 1.96 (as shown in Table~\ref{tab:tbl2}), and are included in all figures.

\begin{deluxetable*}{cclcrrrcr}
  \tablecaption{Observed and derived properties of molecular gas for the target galaxies.}
  \label{tab:tbl2}
  \tablewidth{0pt}
  \tablehead{
    \colhead{Target No.} & \colhead{Obs. No.} & \colhead{Telescope} & 
    \colhead{Useful Exp.}  & 
    \colhead{$I_{\rm CO(1-0)}$} & \colhead{$I_{\rm CO(2-1)}$} & 
    \colhead{$\log(L{'}_{\rm CO(1-0)})$} &
    \colhead{$S_{\rm CO,tot}/S_{\rm CO,obs}$} &
    \colhead{log(\mh)}   \\
    \colhead{} & \colhead{} & \colhead{} & \colhead{(min)} & 
    \colhead{(K km s$^{-1}$)} & \colhead{(K km s$^{-1}$)}  &  
    \colhead{(${\rm K\;km\;s}^{-1}\;{\rm pc^{2}}$)} & \colhead{} & \colhead{(\msolar)}
  }
  \decimalcolnumbers
  \startdata
1&	1	&	PMO	&	55	&	1.75	$\pm$	0.23&		$\dots$		&	9.11	$\pm$	0.06	&	$\dots$	&	9.74	\\
2&	2	&	PMO	&	358	&	1.21	$\pm$	0.15&		$\dots$		&	9.19	$\pm$	0.06	&	$\dots$	&	9.82	\\
&	3	&	JCMT	&	48	&		$\dots$		&	5.36	$\pm$	0.72&	9.27	$\pm$	0.06	&	1.15	&	9.96	\\
3&	4	&	PMO	&	54	&	2.07	$\pm$	0.24&		$\dots$		&	9.56	$\pm$	0.05	&	$\dots$	&	10.19	\\
4&	5	&	JCMT	&	53.3	&		$\dots$		&	2.24	$\pm$	0.58&	8.47	$\pm$	0.11	&	1.11	&	9.15	\\
5&	6	&	PMO	&	167	&	1.64	$\pm$	0.11&		$\dots$		&	8.86	$\pm$	0.03	&	$\dots$	&	9.49	\\
6&	7	&	PMO	&	135	&	1.44	$\pm$	0.18&		$\dots$		&	9.09	$\pm$	0.05	&	$\dots$	&	9.72	\\
&	8	&	CSO	&	6.7	&		$\dots$		&	2.28	$\pm$	0.73&	9.03	$\pm$	0.14	&	1.05	&	9.68	\\
7&	9	&	JCMT	&	66.2	&		$\dots$		&	2.17	$\pm$	0.25&	8.45	$\pm$	0.05	&	1.63	&	9.30	\\
8&	10	&	PMO	&	341	&	1.19	$\pm$	0.16&		$\dots$		&	8.91	$\pm$	0.06	&	$\dots$	&	9.54	\\
&	11	&	JCMT	&	16	&		$\dots$		&	8.18	$\pm$	1.09&	9.19	$\pm$	0.06	&	1.1	&	9.86	\\
9&	12	&	PMO	&	328	&	1.62	$\pm$	0.14&		$\dots$		&	9.11	$\pm$	0.04	&	$\dots$	&	9.74	\\
&	13	&	JCMT	&	57.1	&		$\dots$		&	4.1	$\pm$	0.41&	8.95	$\pm$	0.04	&	1.29	&	9.69	\\
10&	14	&	JCMT	&	80	&		$\dots$		&	1.6	$\pm$	0.24&	8.22	$\pm$	0.07	&	1.65	&	9.07	\\
11&	15	&	JCMT	&	36.3	&		$\dots$		&	2.61	$\pm$	0.68&	8.91	$\pm$	0.11	&	1.96	&	9.84	\\
&	16	&	CSO	&	30	&		$\dots$		&	1.21	$\pm$	0.30&	8.87	$\pm$	0.11	&	1.39	&	9.65	\\
12&	17	&	JCMT	&	16.4	&		$\dots$		&	\textless	3.58&	\textless	9.14	&      	1.14
	&	\textless	9.83	\\
13&	18	&	JCMT	&	60.5	&		$\dots$		&	4.42	$\pm$	0.42&	9.06	$\pm$	0.04	&	1.23	&	9.78	\\
14&	19	&	PMO	&	249	&	1.33	$\pm$	0.14&		$\dots$		&	9.09	$\pm$	0.05	&	$\dots$	&	9.72	\\
15&	20	&	PMO	&	133	&	1.21	$\pm$	0.14&		$\dots$		&	9.02	$\pm$	0.05	&	$\dots$	&	9.65	\\
&	21	&	JCMT	&	40	&		$\dots$		&	4.04	$\pm$	0.47&	8.98	$\pm$	0.05	&	1.17	&	9.68	\\
16&	22	&	JCMT	&	60.5	&		$\dots$		&	2.82	$\pm$	0.86&	8.76	$\pm$	0.13	&	1.7	&	9.62	\\
17&	23	&	JCMT	&	110.6	&		$\dots$		&	3.2	$\pm$	0.39&	8.84	$\pm$	0.05	&	1.39	&	9.62	\\
18&	24	&	JCMT	&	40	&		$\dots$		&	6.3	$\pm$	0.94&	9.15	$\pm$	0.06	&	1.28	&	9.89	\\
19&	25	&	PMO	&	51	&	1.68	$\pm$	0.22&		$\dots$		&	9.19	$\pm$	0.06	&	$\dots$	&	9.82	\\
&	26	&	JCMT	&	68	&		$\dots$		&	3.01	$\pm$	0.44&	8.88	$\pm$	0.06	&	1.27	&	9.62	\\
20&	27	&	JCMT	&	32	&		$\dots$		&	2.95	$\pm$	0.91&	8.77	$\pm$	0.13	&	1.22	&	9.49	\\
21&	28	&	PMO	&	204	&	1.58	$\pm$	0.21&		$\dots$		&	9.12	$\pm$	0.06	&	$\dots$	&	9.75	\\
22&	29	&	PMO	&	207	&	1.66	$\pm$	0.21&		$\dots$		&	9.06	$\pm$	0.06	&	$\dots$	&	9.69	\\
23&	30	&	PMO	&	141	&	2.61	$\pm$	0.18&		$\dots$		&	9.66	$\pm$	0.03	&	$\dots$	&	10.29	\\
&	31	&	JCMT	&	26.3	&		$\dots$		&	8.33	$\pm$	1.32&	9.60	$\pm$	0.07	&	1.14	&	10.29	\\
24&	32	&	PMO	&	130	&	1.93	$\pm$	0.23&		$\dots$		&	9.15	$\pm$	0.05	&	$\dots$	&	9.78	\\
25&	33	&	PMO	&	89	&	2.96	$\pm$	0.26&		$\dots$		&	9.49	$\pm$	0.04	&	$\dots$	&	10.12	\\
26&	34	&	JCMT	&	52.6	&		$\dots$		&	\textless	1.29 &	\textless	8.51	&       1.49
	&	\textless	9.32	 \\
&	35	&	CSO	&	23	&		$\dots$		&	2.11	$\pm$	0.42&	9.02	$\pm$	0.09	&	1.26	&	9.75	\\
27&	36	&	PMO	&	137	&	2.55	$\pm$	0.34&		$\dots$		&	9.22	$\pm$	0.06	&	$\dots$	&	9.85	\\
&	37	&	JCMT	&	16	&		$\dots$		&	4.32	$\pm$	1.18&	8.89	$\pm$	0.12	&	1.4	&	9.67	\\
28&	38	&	PMO	&	132	&	2.67	$\pm$	0.25&		$\dots$		&	9.78	$\pm$	0.04	&	$\dots$	&	10.41	\\
29&	39	&	JCMT	&	26.3	&		$\dots$		&	\textless	1.69&	\textless	8.60	&	1.27	&	\textless	9.34	\\
30&	40	&	JCMT	&	49.4	&		$\dots$		&	\textless	1.68&	\textless	8.72	&	3.27 \tablenotemark{a}	&	\textless	9.35	\\
31&	41	&	JCMT	&	16	&		$\dots$		&	\textless	2.5&	\textless	8.85	&	1.82	&	\textless	9.74	\\
  \enddata
  \tablecomments{From left to right, the columns are: (1) serial number unique to the
    target; (2) serial number of the observation; (3) telescope
    used for the observation; (4) on-source observing time;
    (5)\&(6) CO(1-0) or CO(2-1) integrated intensity (and its uncertainty) of the CO emission
    line; (7) derived CO(1-0) luminosity (and its uncertainty) without aperture correction; (8) aperture corrections for the CO(2-1) observations; (9)molecular gas mass (and its uncertainty) after aperture correction, which is computed as introduced in Section~\ref{sec:comp_mh}.}
\tablenotetext{a} {The $R_{\rm e}$ of this galaxy is unreliable, we show its upper limit of \mh without aperture correction.}
\end{deluxetable*}

\begin{figure*}
\centering \includegraphics[width=0.45\textwidth]{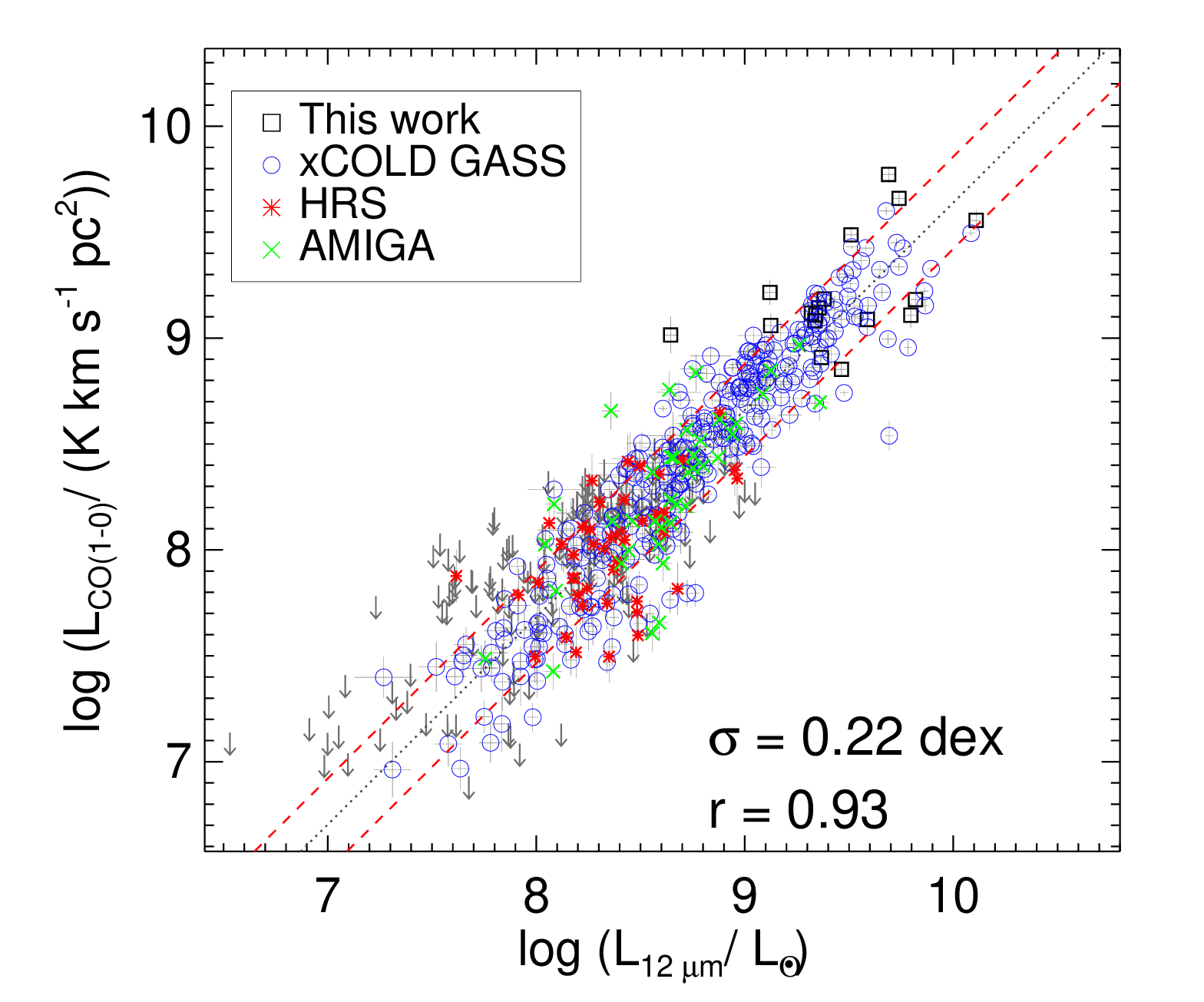}
\includegraphics[width=0.45\textwidth]{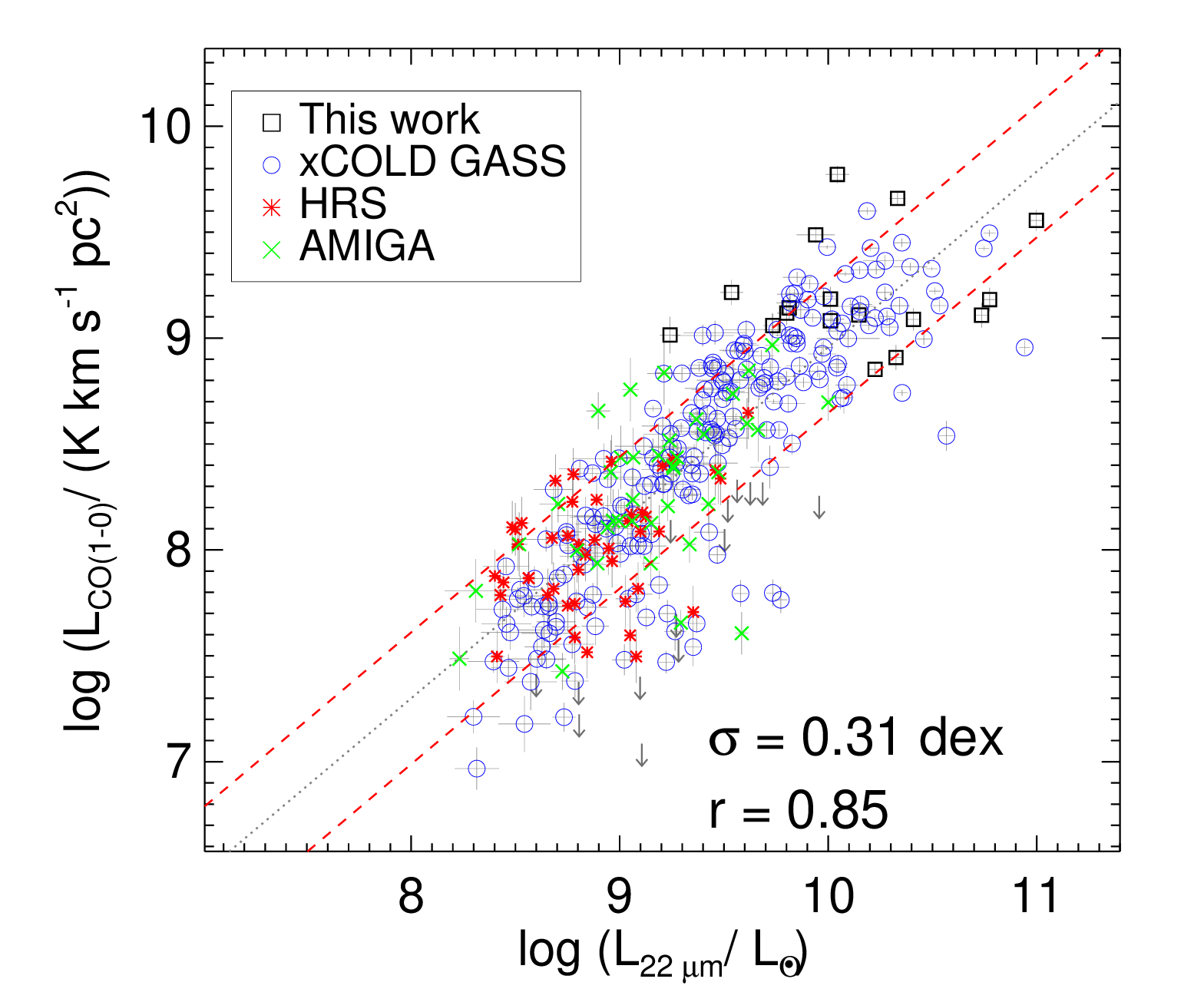}
          		
\caption{Correlation of the CO(1-0) luminosities with the 
  mid-infrared  luminosities from \textit{WISE} measured in the 12 (left
  panel) and  22 \micron\ (right panel) band . Different colors/symbols indicate detections in 
  different CO samples, as indicated in 
  the top-left corner of each panel, and the grey error bars to show their measurement uncertainties, 
 while the dark grey downward-pointing arrows mean upper limits from xCOLD GASS. In each panel, the dotted
  black line and two dashed red lines show respectively the best-fitting
  linear relation (with parameters listed in Table~\ref{tab:tbl3}) and the $1\sigma$ total/observed scatter for detections only obtained from all the samples together. 
  The total/observed scatter ($\sigma$) and the Spearman's correlation coefficient ($r$) for all the detections are listed in the
  bottom-right corner of each panel.} 
\label{fig:co10_w}
\end{figure*}

\begin{figure*}
\centering \includegraphics[width=0.45\textwidth]{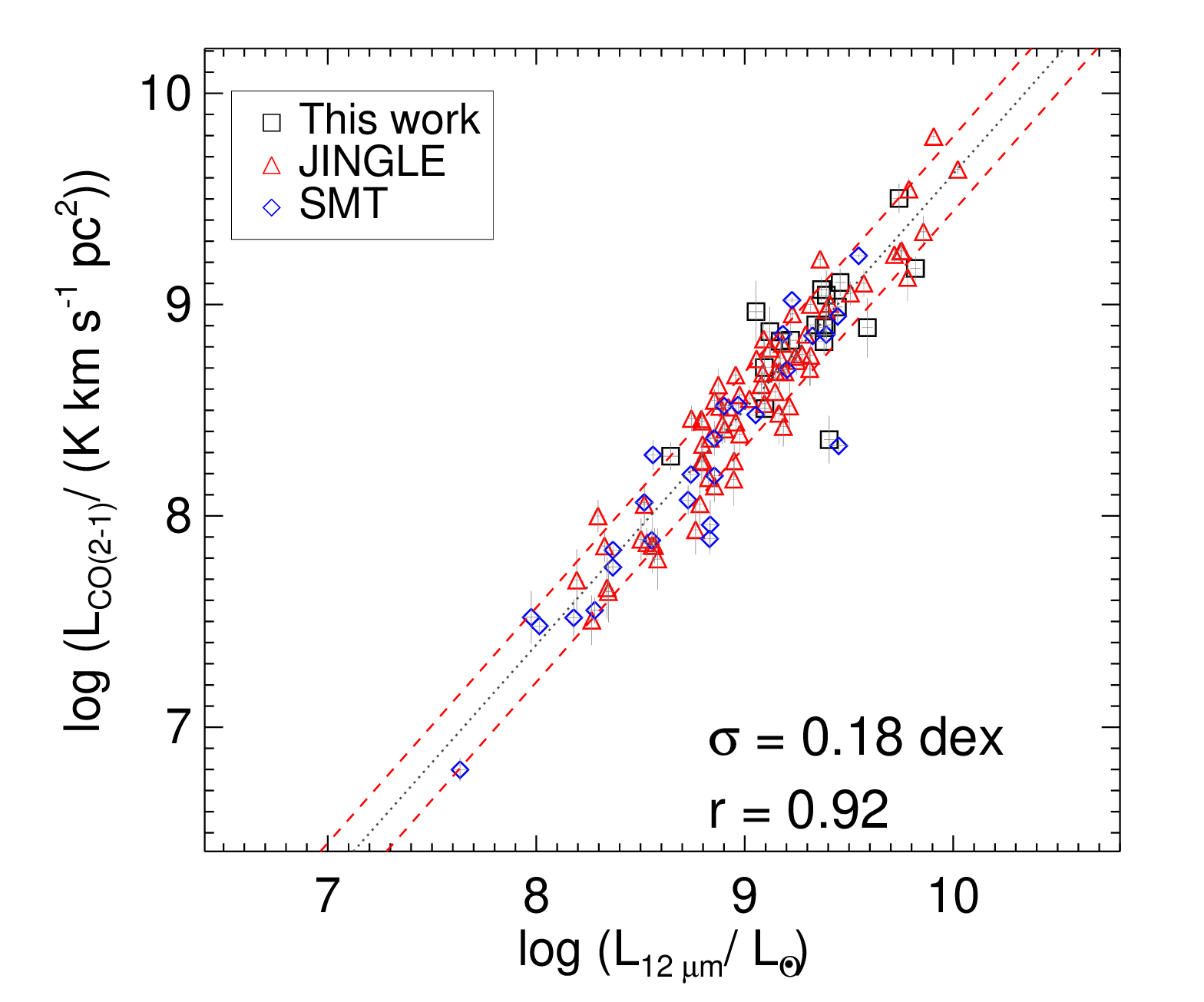}
\includegraphics[width=0.45\textwidth]{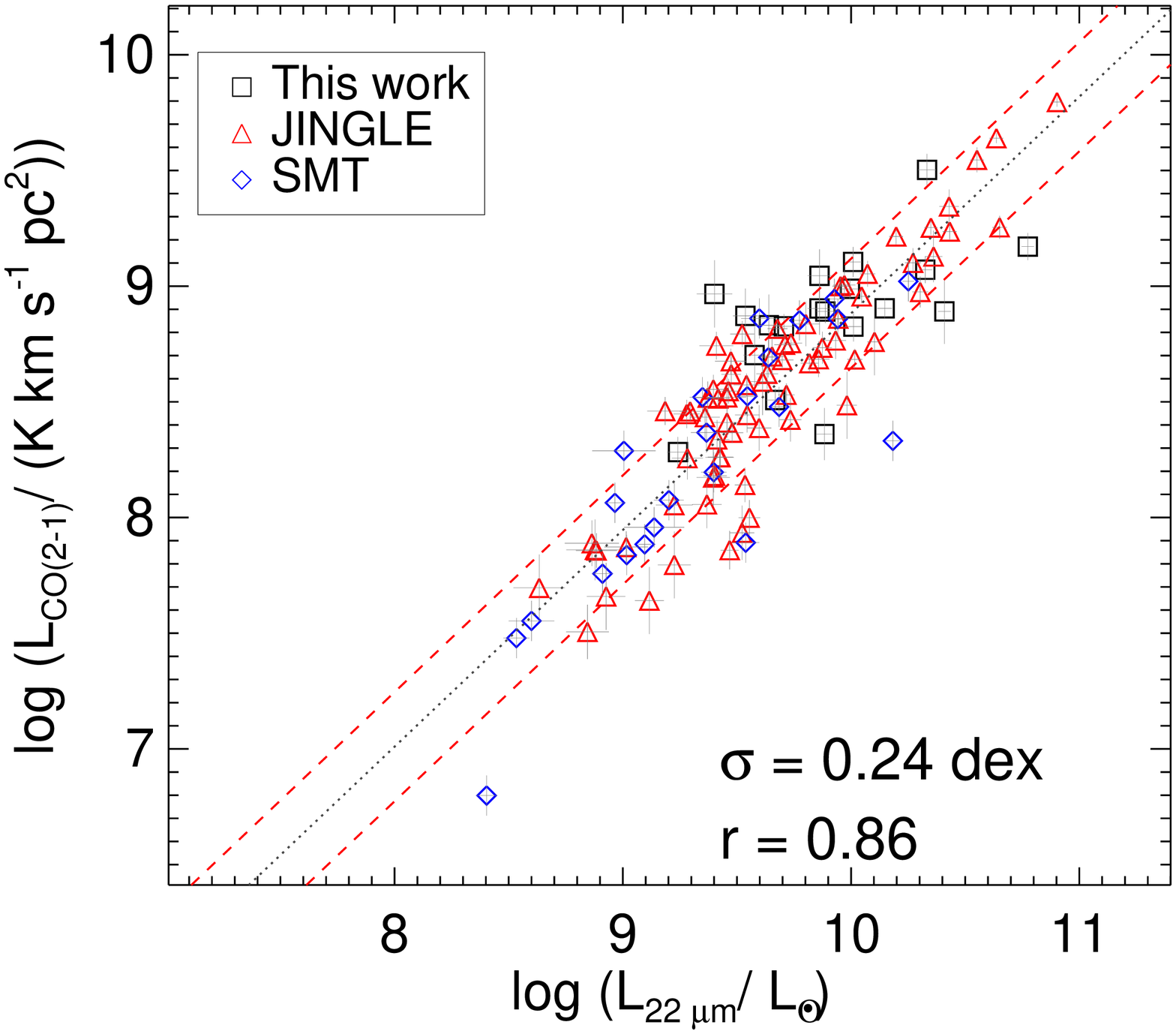}
\caption{As Figure~\ref{fig:co10_w}, but for the CO(2-1) luminosities. }
\label{fig:co21_w}
\end{figure*}

\section{Correlations between the CO and MIR luminosities}
\label{sec:co_mir}

In this section we first examine the correlations of the CO (1-0) and
CO (2-1) luminosities with the mid-infrared luminosities from the
\textit{WISE} 12\micron\ and 22\micron\ bands. We
then extend the correlation of the CO(1-0)  luminosity with the
12 \micron\ luminosity to derive an estimator  of the CO(1-0) luminosity,
by considering other galaxy properties as
additional parameters  to the 12 \micron\ luminosity. Finally we apply this
estimator to our galaxies with only CO(2-1) observations, as well as those
from the JINGLE and SMT surveys, examining the dependence of the
CO(2-1)-to-(1-0) line ratio on a variety of galaxy properties.

We have measured the 12 and 22 \micron\ luminosities of all the
  galaxies to be included in the following analyses. For each galaxy we
reprocessed the W3 (12 \micron) and W4 (22 \micron) images
from \textit{WISE} using the SEXTRACTOR package \citep{Bertin1996}.  We
manually adjusted the shape and size of the ellipse to properly cover
the MIR emission of each galaxy. Before estimating the fluxes and
luminosities, we carefully masked out by hand
foreground and background sources as well as neighboring galaxies
identified in the corresponding SDSS image. We used redshift from SDSS when computing the
distance.  The 12 and 22 {\micron} luminosities (and their respective uncertainties) thus estimated for our
galaxies are listed in Table~\ref{tab:tbl1}, with median uncertainties of 0.021 and 0.026 dex, respectively.
The uncertainty includes both the photometric uncertainty (is consistent with flux uncertainty in \textit{WISE} catalog) 
and the uncertainty of the magnitude zero-point \citep{Jarrett2011}. 

\subsection{Correlations between CO and mid-infrared luminosities}
\label{sec:co_MIR}
Figure~\ref{fig:co10_w} shows the correlation of the CO(1-0) luminosities,
$L_{\rm CO(1-0)}$, with the mid-infrared luminosities measured from \textit{WISE}
in the 12 (left panel) and 22 \micron\  (right panel) band. Our
galaxies observed with the PMO 13.7-m telescope are plotted as black
squares. For comparison, we show the detected galaxies from xCOLD
GASS \citep{Saintonge2017}, HRS \citep{Boselli2014} and
AMIGA \citep{Lisenfeld2011}, that have significant detections in
both CO(1-0) and \textit{WISE} ($S/N$ > 3), as blue circles, red stars and green crosses, respectively. 
The figure thus includes all types of galaxies, early- and late-type galaxies as well as
active galactic nucleus (AGN) hosts, interacting/paired galaxies and luminous infrared galaxies
(LIRGs). Analogously, Figure~\ref{fig:co21_w} shows the correlation of the CO(2-1)
luminosities, $L_{\rm CO(2-1)}$, with the \textit{WISE} 12 and
22 \micron\  luminosities. Our galaxies detected with JCMT and/or CSO are
plotted as black squares, while those detections from JINGLE \citep{Saintonge2018} and the
SMT observations of \citet{Jiang2015} are plotted  as red triangles
and blue diamonds, respectively. 

\begin{table*}\footnotesize
        \centering
	\caption{Best-fit relations between CO and MIR luminosities.}
	\label{tab:tbl3}
	\begin{tabular}{ccccccc}
           
	\hline\hline 
 ${\rm log} (L_{\mbox{12\micron}}{\rm L_\odot})$ &	${\rm log} (L_{\rm CO}/[{\rm K\;km\;s}^{-1}\;{\rm pc^{2}}])$	&	Sample	&	k			&	b			&	$\sigma_{\rm int}$	& corresponding panel	\\\hline
12 \micron	&	CO(1-0) 	&	CO detections (412)	&	0.98	$\pm$	0.02	&	-0.14	$\pm$	0.18	&	0.20	& the left panel of Figure~\ref{fig:co10_w}	\\
12 \micron	&	CO(1-0)		&	CO detections and upper limits (565)	&	1.03	$\pm$	0.02	&	-0.64	$\pm$	0.18	&	0.21	&	\\
12 \micron	&	CO(2-1)		&	CO detections (118)	&	1.11	$\pm$	0.04	&	-1.52	$\pm$	0.33	&	0.15	& the left panel of Figure~\ref{fig:co21_w}	\\

22 \micron	&	CO(1-0)		&	CO detections (318)	&	0.83	$\pm$	0.03	&	0.63	$\pm$	0.30	&	0.30	& the right panel of Figure~\ref{fig:co10_w}	\\
22 \micron	&	CO(1-0)		&	CO detections and upper limits (332)	&	0.84	$\pm$	0.03	&	0.50	$\pm$	0.32	&	0.33	&	\\
22 \micron	&	CO(2-1)		&	CO detections (114)	&	0.94	$\pm$	0.04	&	-0.47	$\pm$	0.44	&	0.21	& the right panel of Figure~\ref{fig:co21_w}	\\
 \hline
	\end{tabular}
	\begin{flushleft}
	\textbf{Notes.} 
	{The relations are parametrized as $y = k x + b$, with all the quantities given in this table. The number of galaxies included in the fitting is indicated in parenthesis in the sample column. The derived intrinsic scatter of each relation is listed as $\sigma_{\rm int}$.}
         \end{flushleft}
\end{table*}

As one can see from Figure~\ref{fig:co10_w}, our galaxies are located
in the upper-right corner of both panels, with the highest CO(1-0)
and MIR luminosities, slightly extending
the trend defined by existing observations towards high luminosities. This is
expected  as our PMO sample was selected to be brightest in the
12 \micron\ band. In Figure~\ref{fig:co21_w}, our galaxies appear to span
a similar range of CO(2-1) luminosities as the JINGLE sample galaxies at mid-to-high end, although
with less coverage at both the high- and low-luminosity ends due to the
much smaller sample size. This is again understandable, as we selected
our JCMT targets randomly from the parent sample. 

From Figures \ref{fig:co10_w} and \ref{fig:co21_w}, we see that the CO
and  MIR luminosities are well-correlated in all cases, but the
correlations are tighter and more linear when the MIR luminosity is
measured in the 12 \micron\ band rather than the 22
\micron\ band. This  is true for both the CO(1-0) and CO(2-1)
lines. To quantify this effect, we performed a Bayesian linear
  regression of $L_{\rm CO}$ as a function of MIR luminosity at both
  12\micron\ and 22\micron, taking into account  uncertainties in both
  the x and y axes using \textit{LinMix} \citep{Kelly2007}, implemented in
the IDL script \textit{linmix\_err.pro}\footnote{Available from the
  NASA IDL Astronomy User's Library
  \url{https://idlastro.gsfc.nasa.gov/ftp/pro/math/linmix_err.pro}}.
In each panel of Figures~\ref{fig:co10_w} and
  \ref{fig:co21_w}, we plot the best-fit line to \textit{detections
  only} and the $1\sigma$ scatter (i.e. the standard deviation of the
data points around the fit) about each line.  The scatters are $\sigma=$ 0.22,
0.31, 0.18 and 0.24 dex, and the Spearman's correlation coefficients
are 0.93, 0.85, 0.92 and 0.86, respectively,
for the correlation of  $L_{\rm CO(1-0)}$ vs. $L_{\mbox{12\micron}}$,
$L_{\rm CO(1-0)}$ vs. $L_{\mbox{22\micron}}$,  $L_{\rm CO(2-1)}$ vs.
$L_{\mbox{12\micron}}$ and $L_{\rm CO(2-1)}$
vs. $L_{\mbox{22\micron}}$, as indicated in each panel. The parameters
of the best-fitting relations are listed  in Table~\ref{tab:tbl3}
including the derived intrinsic scatters.  For the relations with
$L_{\rm CO(1-0)}$, we also carried out fits taking into
  account non-detections in xCOLD GASS (as "censored" data), and the
fitting results do not change significantly. 

The fits suggest that the MIR luminosities are slightly more tightly
correlated with the CO(2-1) luminosities than with the CO(1-0)
luminosities. This can be easily understood, as CO(2-1) is associated
with denser and/or warmer gas than CO(1-0), and is thus  more likely
to be associated with the star-formation traced by the MIR
luminosities.  As the CO(1-0) observations come from several different
telescopes/surveys, the slightly larger scatters in the correlations
of the CO(1-0) observations could also be partly (if not fully)
attributed to the systematic differences between the different
samples.   In particular, the CO(2-1) observations in
Figure~\ref{fig:co21_w} are dominated by the JINGLE sample. For the
CO(1-0) observations, the total/observed (intrinsic) scatter is
reduced to 0.20 (0.18) dex if we consider only the  xCOLD GASS sample.
The K-correction of the MIR luminosities probably also partly
contributes to the scatter in these relations \citep[see ][]{Lee2013}.

\begin{figure}
  \begin{center}
    \includegraphics[width=0.49\textwidth]{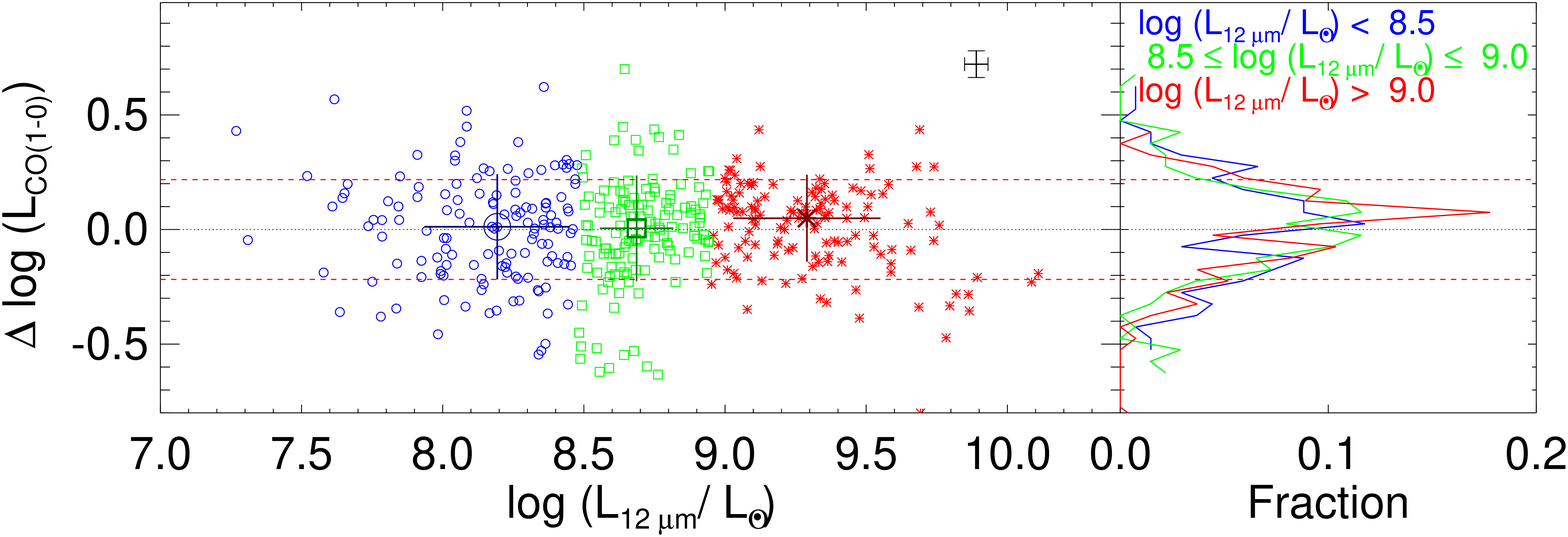}
    \includegraphics[width=0.49\textwidth]{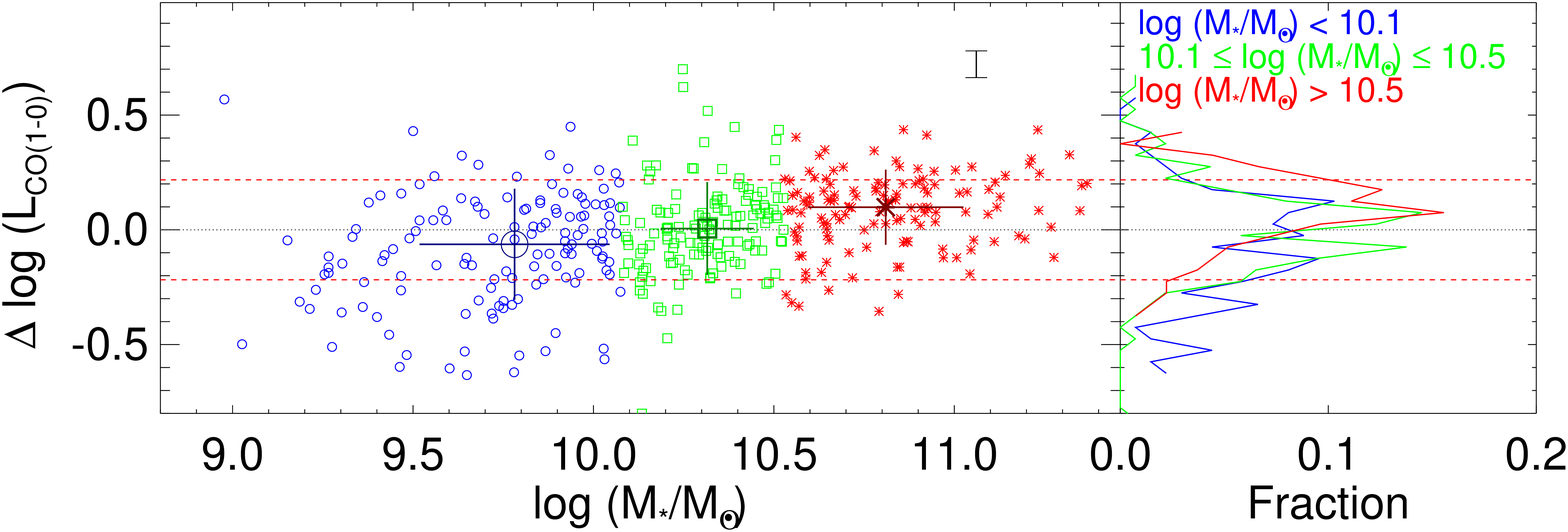}
    \includegraphics[width=0.49\textwidth]{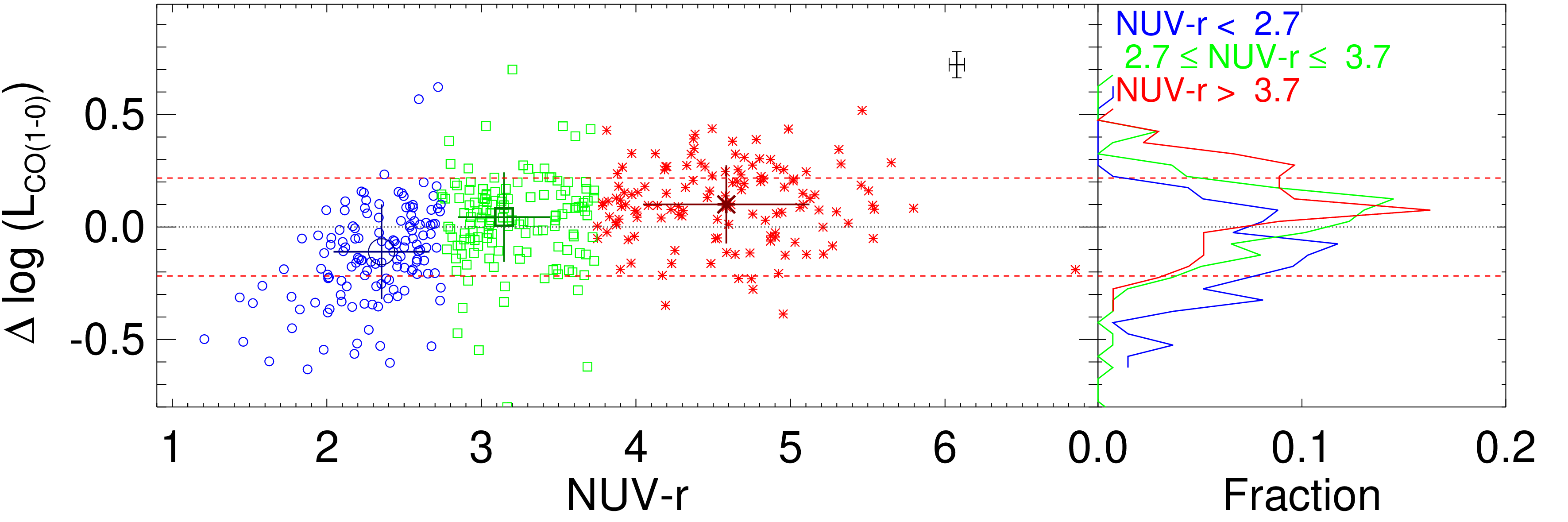}
    \includegraphics[width=0.49\textwidth]{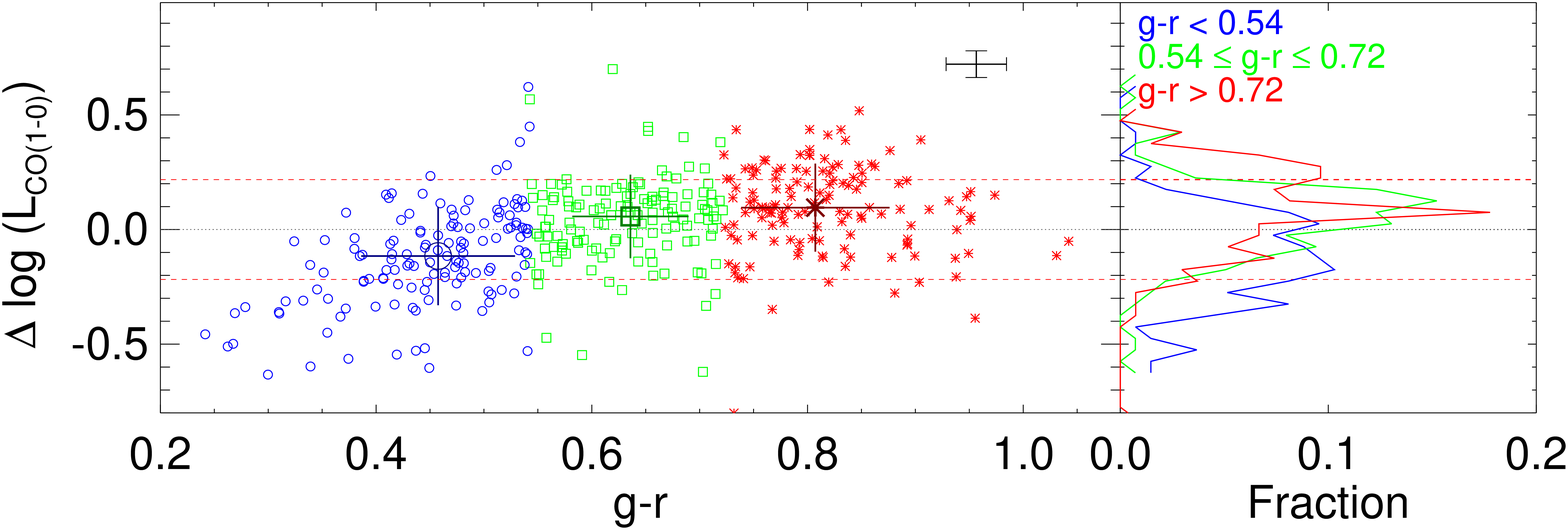}
    \includegraphics[width=0.49\textwidth]{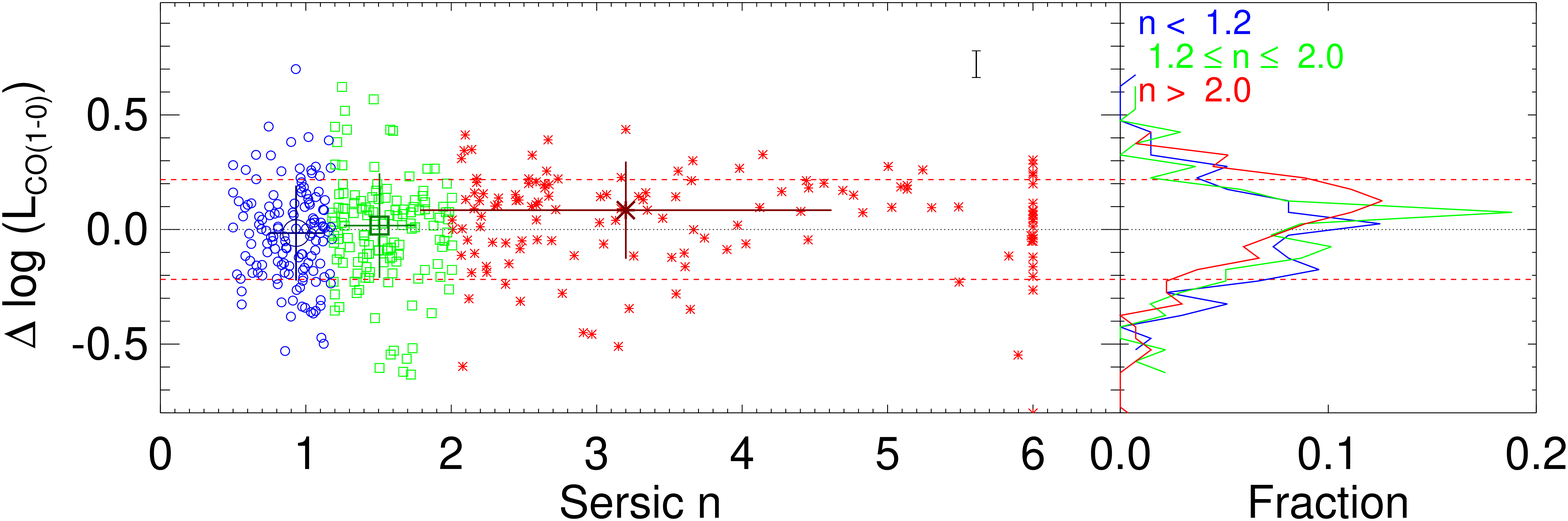}
  \end{center}
  \caption{In the left-hand panels, for CO(1-0) detections, we plot the residuals (${\Delta}$ log($L_{\rm CO(1-0)}) \equiv$ log($L_{\rm CO(1-0),obs}$)- log($L_{\rm CO(1-0),est}$)) of CO(1-0) luminosity predicted from the best-fitting relation (Eq.~\ref{co_12}) between $L_{\rm CO(1-0)}$ and
    $L_{12 \micron}$ and the observed $L_{\rm CO(1-0)}$ as a function of 12 \micron\ luminosity $L_{12 \micron}$, stellar
    mass \mstar\, {\nuvr} color, $g-r$ color and Sersic index $n$ (from top to bottom). 
    The dashed red horizontal lines show the $1\sigma$ total/observed scatter of all galaxies, and 
    the characteristic error bars illustrate the median value of available measurement uncertainties.
    In each panel, we divide the galaxies into 3 sub-samples (with the same number of galaxies in each) according to the parameter considered. 
    Blue circles, green squares and red stars correspond to respectively the smallest,
    intermediate and largest parameter values, and the large symbols show the median and scatter in each bin. In the right-hand panels, the different colors show the
    distribution of the residuals in each sub-sample.  }
  \label{fig:res_est0}
\end{figure}

\begin{figure}
  \begin{center}
    \includegraphics[width=0.49\textwidth]{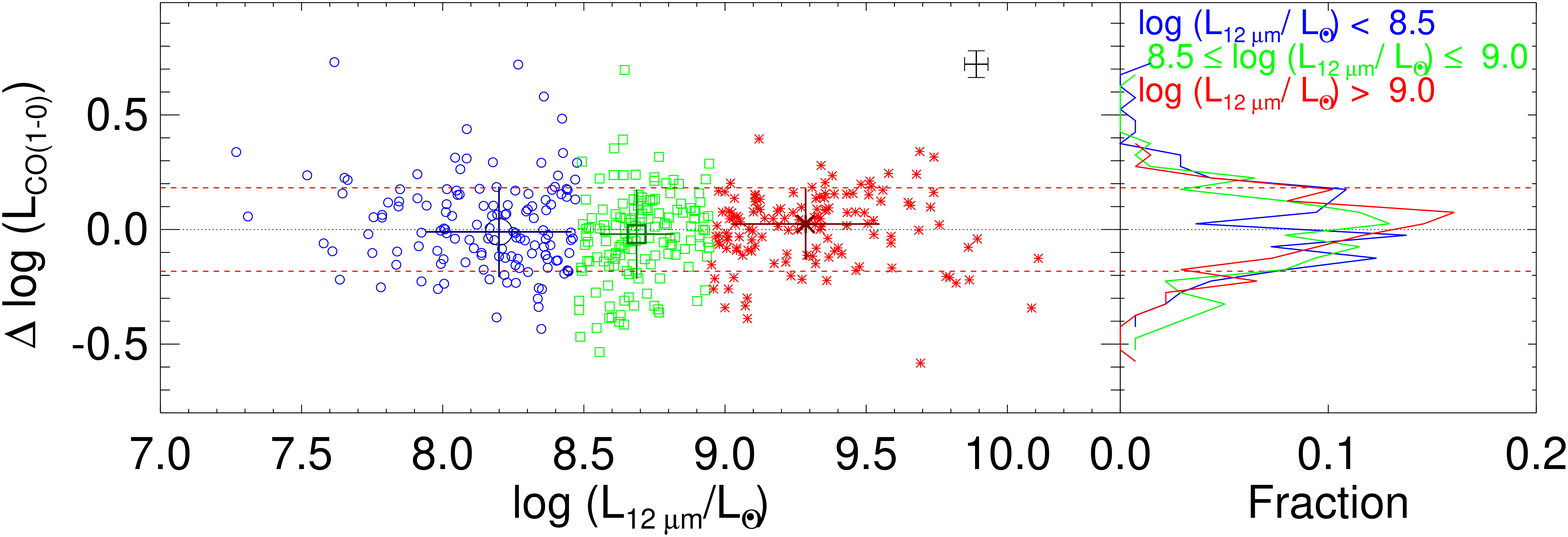}
    \includegraphics[width=0.49\textwidth]{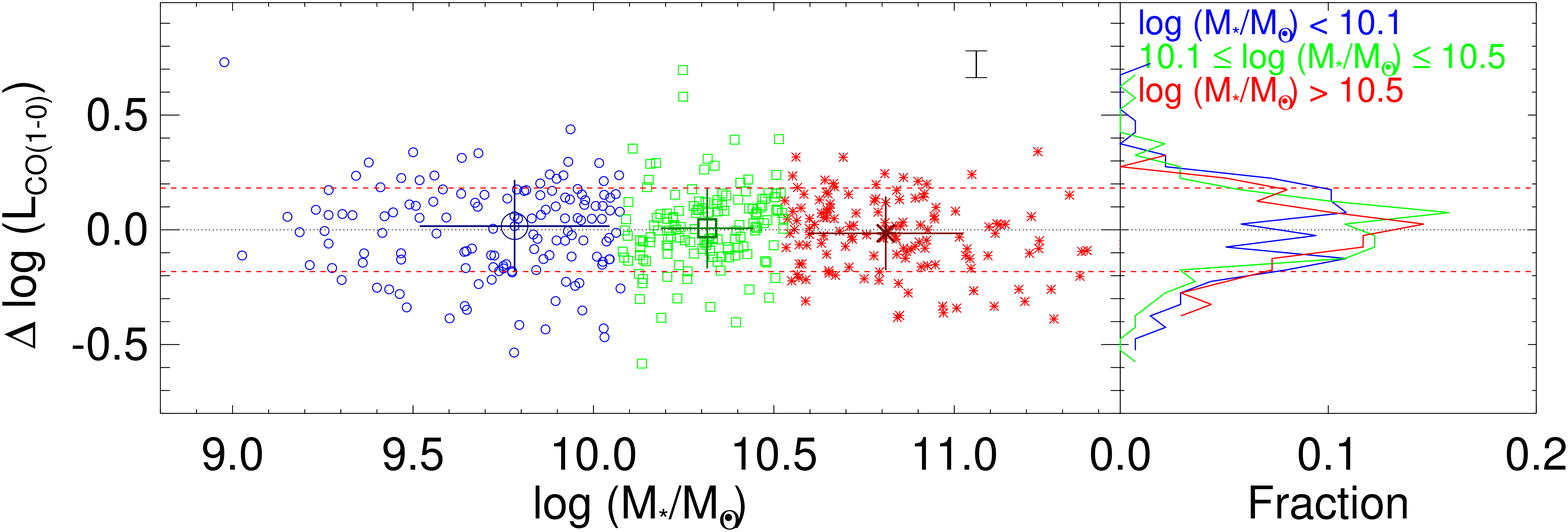}
    \includegraphics[width=0.49\textwidth]{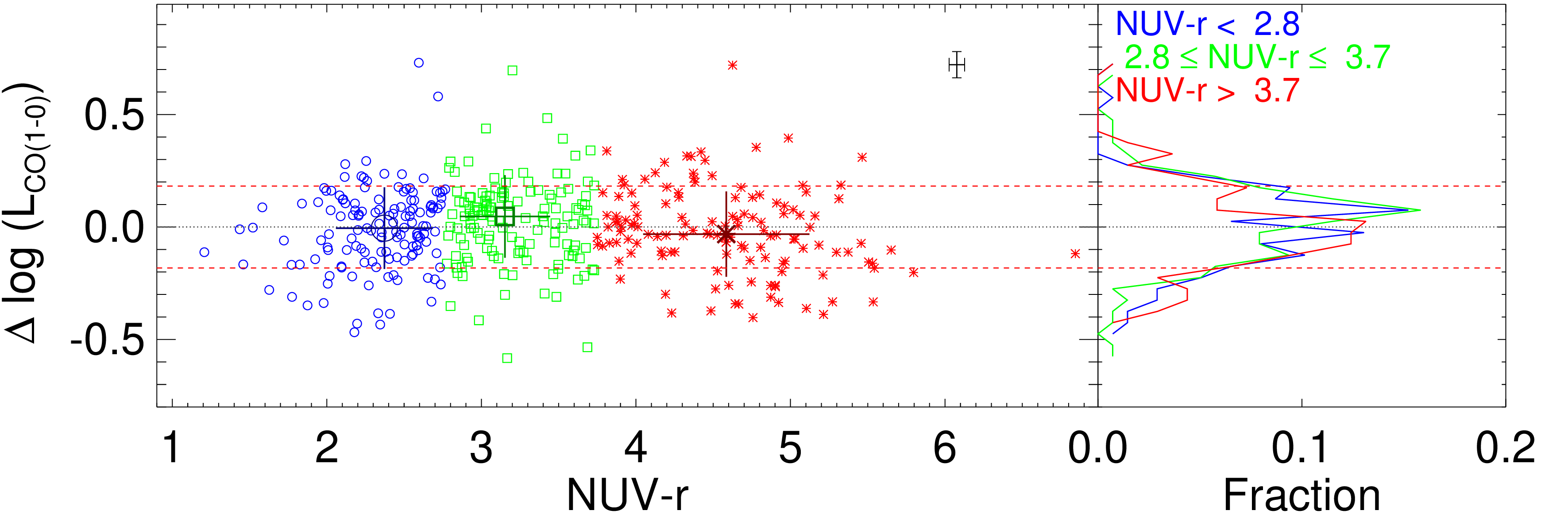}
    \includegraphics[width=0.49\textwidth]{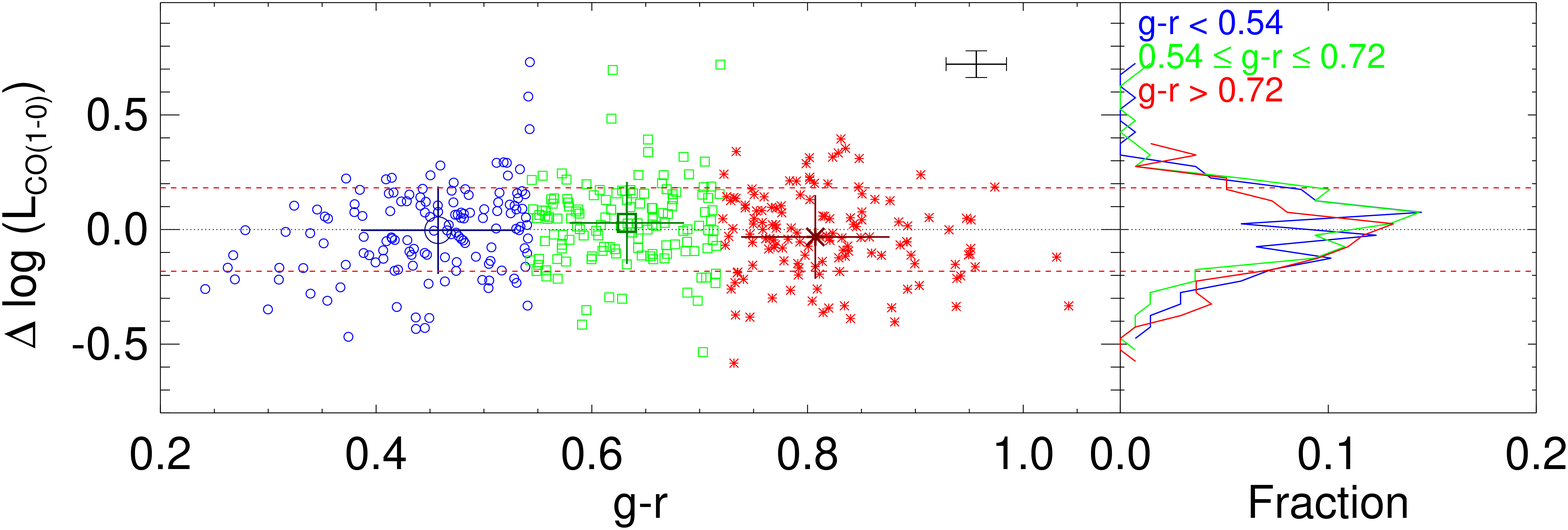}
    \includegraphics[width=0.49\textwidth]{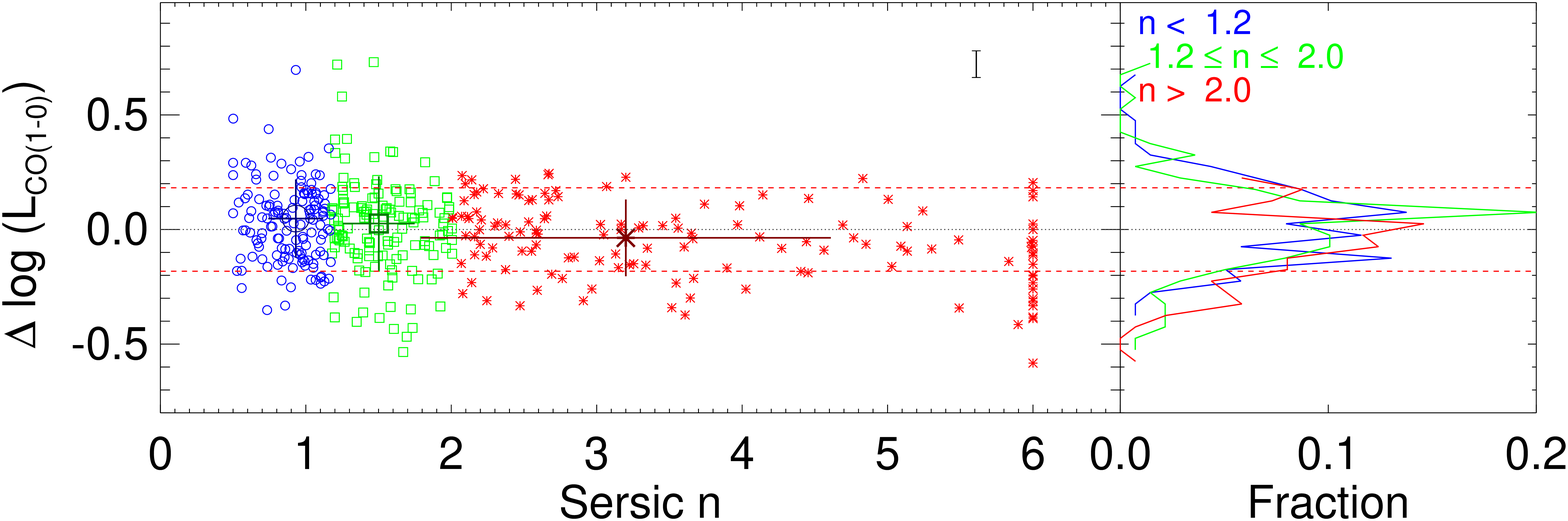}
  \end{center}
  \caption{ As Figure~\ref{fig:res_est0} but for the residuals of the predicted
    value of $L_{\rm CO(1-0)}$ using the three-parameter estimator (Eq.~\ref{co_estim}).  }
  \label{fig:res_est}
\end{figure}

On the other hand, we notice that another more important reason for
the larger total/observed and intrinsic scatters in the $L_{\rm CO(1-0)}$ correlations is the systematic deviation of some of
the low-luminosity galaxies,  many being well below the linear relation
of the whole sample.  As can be clearly seen in
Figure~\ref{fig:co10_w},  this effect occurs mainly at
$L_{\mbox{12\micron}}\lesssim 10^{8}$ \Lsolar and $L_{\mbox{22\micron}}\lesssim
10^{9}$ \Lsolar.  The CO(2-1) samples contain very few
galaxies at these luminosities. When data become
available in the future, it would thus be interesting to
see whether the $L_{\rm CO(2-1)}$ versus  MIR luminosity correlations also show
similar downturns at the low-luminosity end.

\begin{figure*}
\begin{center}
\includegraphics[width=0.45\textwidth]{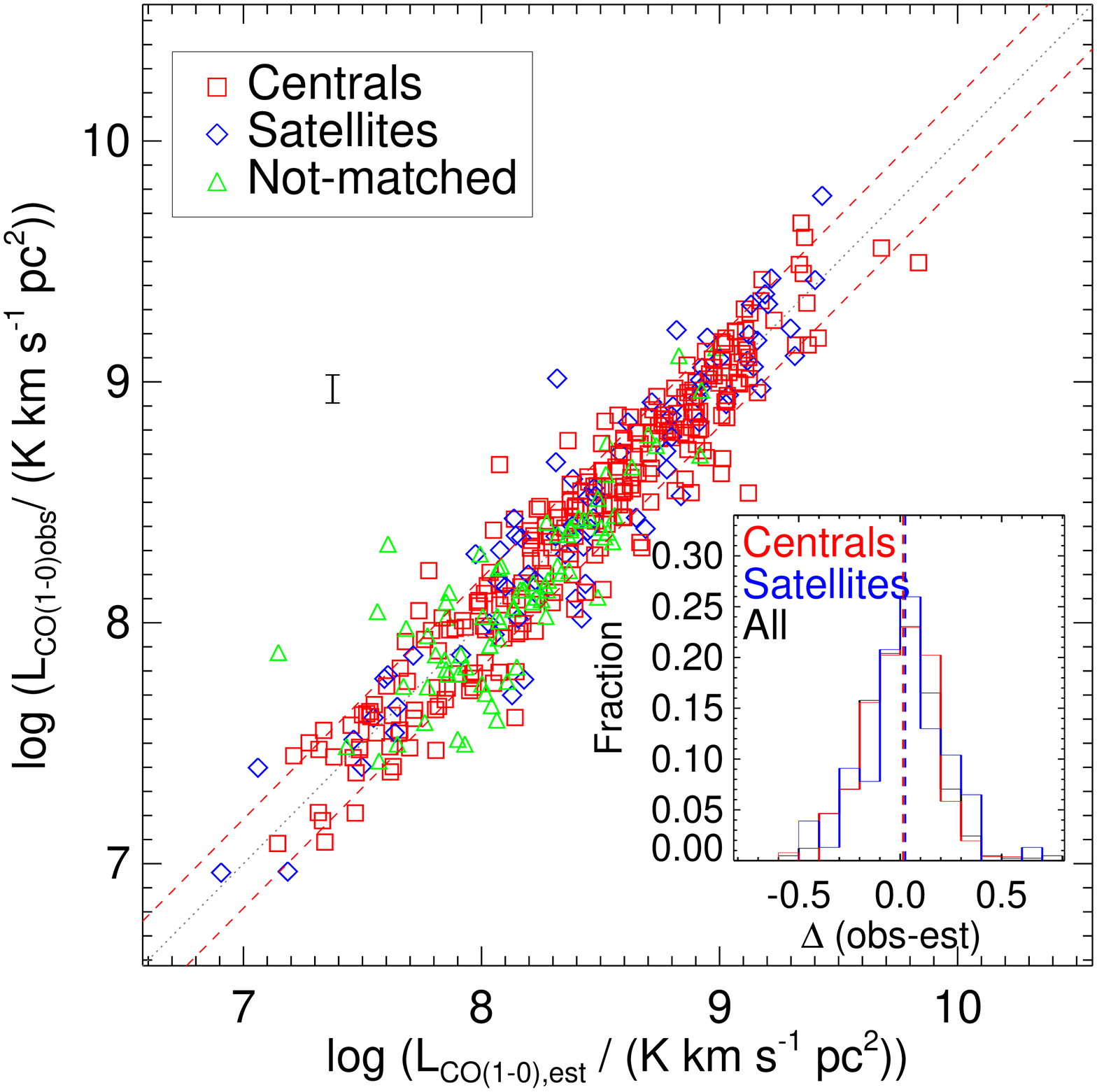}
\includegraphics[width=0.45\textwidth]{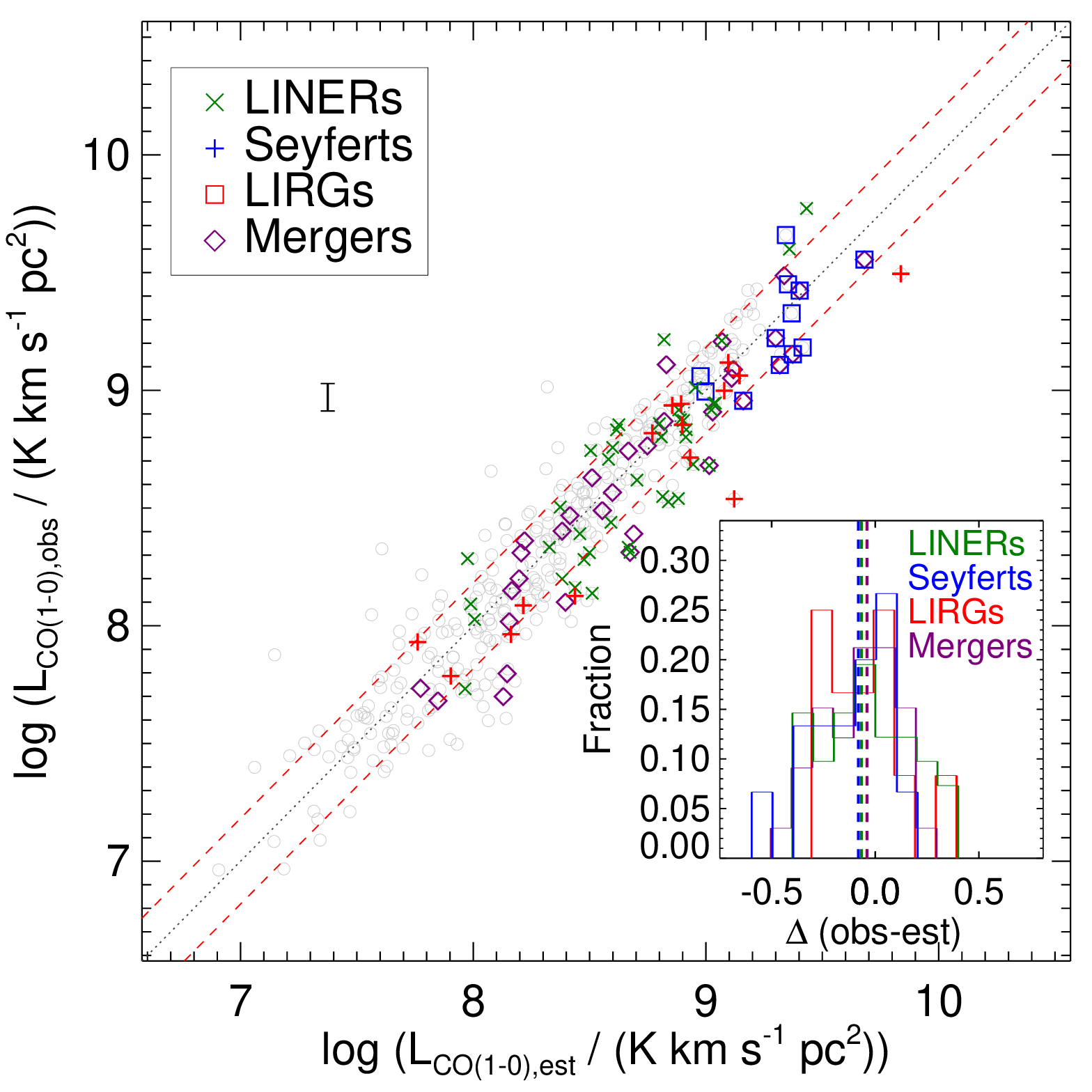}
\end{center}
\caption{ New estimator of the CO(1-0) luminosity ($L_{\rm CO(1-0)}$) 
  based on $L_{12 \micron}$, \mstar\ and $g-r$, as
  well as the distribution of the residuals of some sub-samples.
  In each panel, the dotted black line and two dashed red lines show respectively 
  the best-fitting relation (with parameters listed in Table~\ref{tab:tbl4}) and the $1\sigma$ total/observed scatter (0.18 dex) of all galaxies.
  In the left panel, different symbols/colors are used to highlight any
  difference between central and satellite galaxies.  In the right
  panel, we highlight particular galaxy populations including
  BPT-selected AGN hosts (LINERs as dark green crosses and Seyferts as blue plus signs), 
  LIRGs (red squares) and interacting galaxies or
  mergers (purple diamonds).  In insets, the histograms and vertical
  dashed lines show respectively the distributions and median values of the residuals for
  different sub-samples. }
\label{fig:cen_sat_agn}
\end{figure*}

\subsection{Residuals in the CO vs. MIR luminosity relation}

To better understand the scatters and the systematic deviations
discussed above (Section~\ref{sec:co_MIR}) in relation to the CO
versus 12 \micron\ luminosity relations,  we hereby examine the
residuals about the best-fitting relations as a function of a variety
of galaxy parameters. As a byproduct, this analysis is expected to
produce a new estimator of CO(1-0) luminosity, that is a
linear combination of multiple parameters, thus providing estimated
CO(1-0) luminosities with smaller uncertainties and biases
than those estimated from the 12 {\micron} luminosity alone.

We primarily consider the linear fit between $L_{\rm CO(1-0)}$ and
$L_{\mbox{12\micron}}$, that is 
\begin{footnotesize}
\begin{equation} 
\label{co_12}
\textrm{log} \Big(\frac{L_{\rm CO(1-0)}}{{\rm K\;km\;s}^{-1}\;{\rm pc^{2}}} \Big) = (0.98 \pm 0.02) \textrm{log} \Big(\frac{L_{12 \micron}}{\rm L_\odot}\Big)-(0.14 \pm 0.18),
\end{equation}
\end{footnotesize}
as indicated in the left panel of Figure~\ref{fig:co10_w}.  In
Figure~\ref{fig:res_est0} we plot the residual of the CO(1-0)
luminosity as a function of five different galaxy parameters  (from
top to bottom) for all available galaxies: 12 \micron-band luminosity
($L_{\mbox{12\micron}}$), stellar mass ($M_\ast$), $NUV-r$ color,
$g-r$ color and Sersic index ($n$).  The residual is defined as the
logarithm of the ratio of the observed CO(1-0) luminosity to
the estimated one. In this case, a positive (negative) residual
indicates a higher (lower) molecular gas mass than that predicted by
$L_{\mbox{12\micron}}$.  Optical colors and the $NUV-r$ color are
known to be sensitive to the cold gas fraction of galaxies
\citep{Saintonge2011}, while the Sersic index is a structural
parameter whereby larger $n$ indicate earlier-type morphologies (and a
prominent bulge for late-type galaxies). In each panel, galaxies in
different bins of the parameter considered are plotted with different
symbols/colors. The two red dotted horizontal lines in each panel
indicate the $1\sigma$ scatter of all the sample galaxies about the
best-fitting relation,  $0.22$ dex in this case.  To the right of each
panel we add a smaller panel showing the histogram of the $L_{\rm
  CO(1-0)}$  residuals for the three sub-samples including the
  same number of objects defined by the  parameter considered.

Overall, the residuals are constant about zero with no/weak dependence
for $L_{\mbox{12\micron}}$ (as expected) and the Sersic index $n$, but
are significantly negative for galaxies with the lowest masses
(${M_\ast} \lesssim 10^{10}$ \msolar) and bluest colors
(\nuvr\ $\lesssim 3$ and/or $g-r \lesssim 0.5$). This echoes the
downturn at the low-luminosity end seen above in the $L_{\rm
  CO(1-0)}-L_{\mbox{12\micron}}$ relation (Figure~\ref{fig:co10_w}).



\begin{figure*}
\centering
  \includegraphics[width=0.31\textwidth,clip=true]{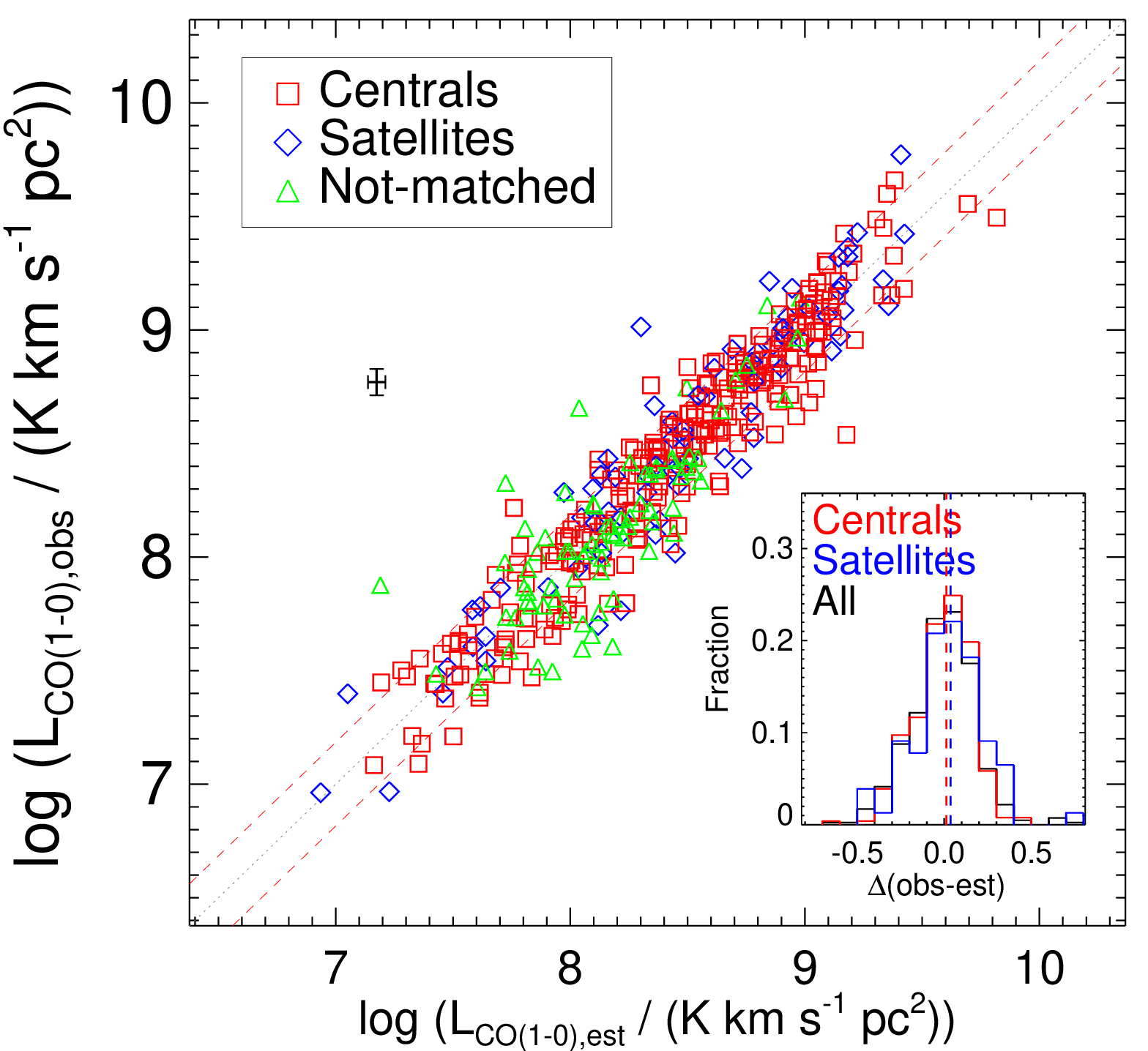}
\includegraphics[width=0.31\textwidth,clip=true]{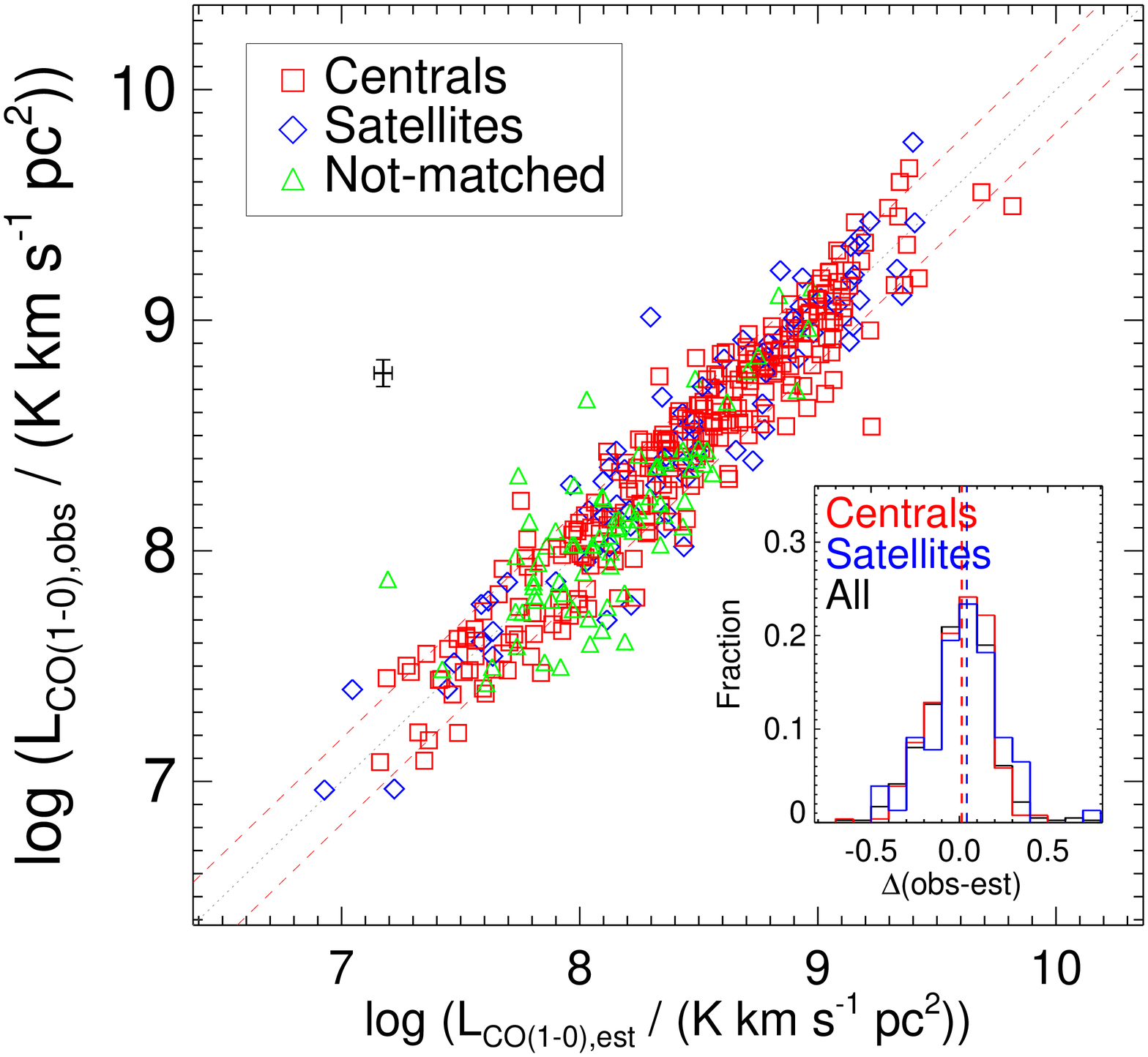}
\includegraphics[width=0.31\textwidth,clip=true]{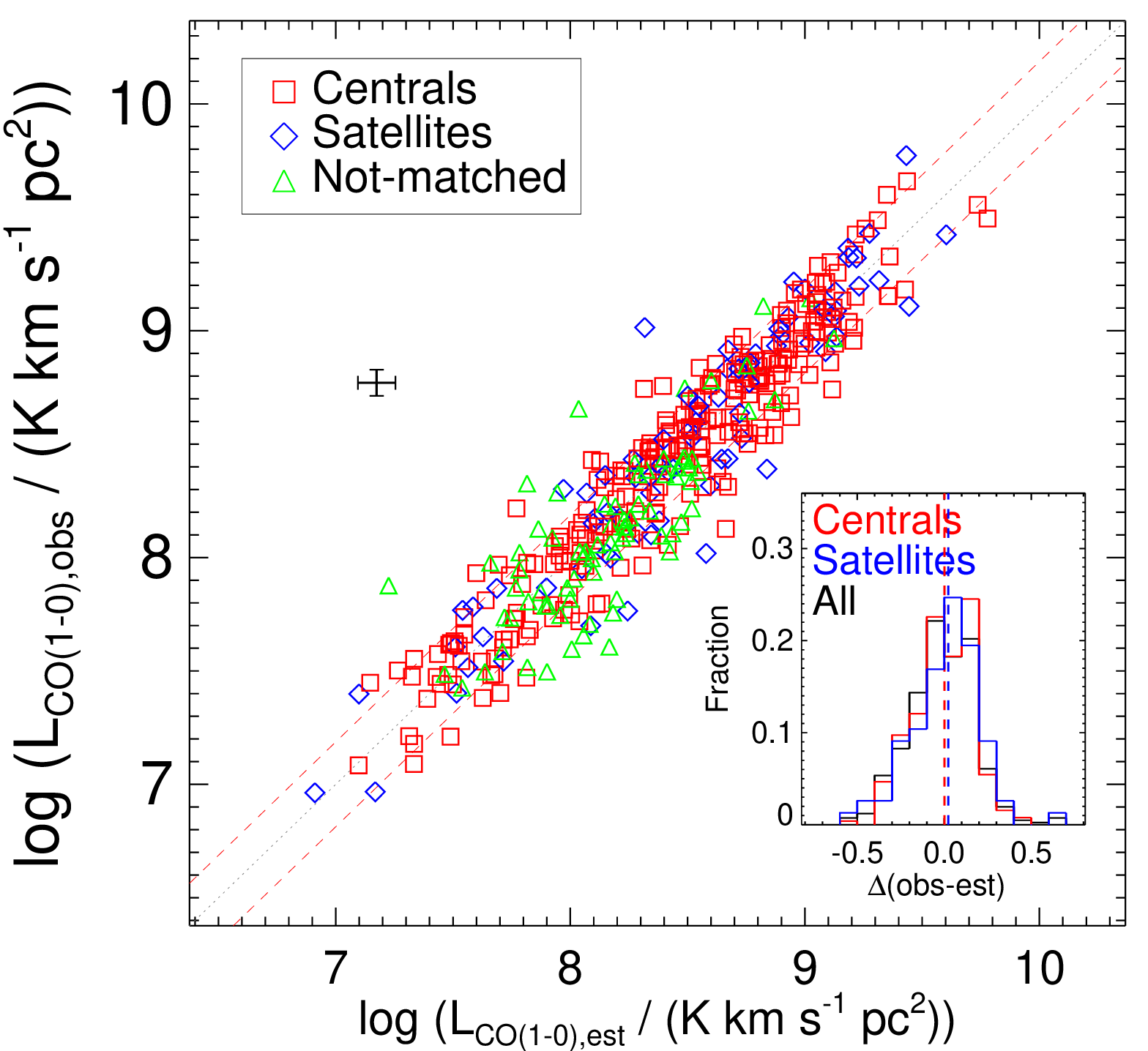}
\caption{ As Figure~\ref{fig:cen_sat_agn} but for alternative estimators of the CO(1-0) luminosity ($L_{\rm CO(1-0)}$)
  based on $L_{12 \micron}$, $g-r$ color and $r$--band
  luminosity (left); $L_{12 \micron}$, $g-r$ color and $z$-- band luminosity (center); $L_{12 \micron}$ , $r$--and $z$--band luminosities (right). The characteristic error bars shown illustrate the median (direct or derived) measurement uncertainties. }
\label{fig:oth_est}
\end{figure*}

\subsection{A three-parameter estimator of CO(1-0) luminosity}
\label{sec:3para}
We therefore include the stellar mass and $g-r$ color as additional
parameters, and estimate $L_{\rm CO(1-0)}$ using
\textit{LinMix} (see above) over multiple parameters 
($L_{\rm 12\micron}$, $M_\ast$ and $g-r$) as provided in Table~\ref{tab:tbl4}:
\begin{footnotesize}
\begin{equation} 
\begin{split}
\label{co_estim}
\textrm{log} \Big(\frac{L_{\rm CO(1-0)}}{{\rm K\;km\;s}^{-1}\;{\rm pc^{2}}} \Big) = (0.76 \pm 0.03) \textrm{log} \Big(\frac{L_{12 \micron}}{\rm L_\odot}\Big) + (0.29 \pm 0.04) (g-r)\\ + (0.29 \pm 0.08) \textrm{log} \Big(\frac{M_\ast}{\rm M_\odot} \Big) - (1.41 \pm 0.25).
\end{split}
\end{equation}
\end{footnotesize}

Figure~\ref{fig:res_est} shows the residuals of the CO(1-0)
luminosities predicted by this 3-parameter estimator as a function of
$L_{\mbox{12\micron}}$, $M_\ast$, \nuvr, $g-r$ and Sersic index $n$,
in the same manner as  Figure~\ref{fig:res_est0}.  The residuals show
no dependence on any parameter.  The additional two parameters also
improve a bit the estimate for the high-mass or red galaxies, though
which are used to mainly remove the residuals in low mass and blue
galaxies.  The total/observed scatter of all the galaxies about the
best-fitting relation is  0.18 dex, considerably smaller than the
scatter of 0.22 dex when estimating $L_{\rm CO(1-0)}$ from the
12\micron\ luminosity alone.

One might wonder whether it is necessary to include both $M_\ast$ and
$g-r$ as additional parameters.  In fact, we have attempted to add
only one parameter, either $M_\ast$ or \nuvr\ or $g-r$, in addition to
the 12\micron\ luminosity, and we find none of them taken alone can
provide unbiased estimates. A third parameter is always necessary to
remove systematic biases, although it does not help to significantly
reduce the scatter.   We have also examined a 3-parameter estimator
combining $L_{\mbox{12\micron}}$, $M_\ast$ and \nuvr, i.e. replacing
$g-r$ by \nuvr\ in Eq.~\ref{co_estim}. We find the estimator using
$g-r$ works better, yielding smaller biases and slightly smaller
scatter.

One may also worry about possible biases in the CO(1-0) luminosities
of galaxies located in different environments, that were previously
shown to  influence the cold gas content of galaxies
(e.g. \citealp{Li2012}; \citealp{Zhang2013}).  In the left panel of
Figure~\ref{fig:cen_sat_agn}, we compare the observed CO luminosities
of the galaxies in this work with those predicted by our estimator,
showing the results for central galaxies (red squares) and satellite
galaxies (blue diamonds) differently (unmatched galaxies are shown as
green triangles).  The central/satellite classification is taken from
the SDSS galaxy group catalog constructed by \citet{Yang2007}
in which the central galaxy of a given group of galaxies
  is defined to be the most massive galaxy in the group.
The inset shows the histograms of the residuals for both
central and satellite galaxy subsets; there is no obvious bias.  We
also carry out a Kolmogorov-Smirnov (K$-$S) test between the residuals
of the centrals/satellites and those of the total sample.  The K$-$S
probability (p-value) is 0.98 and 0.41, respectively, suggesting
the two samples to be drawn from the same parent sample randomly.

In the right-hand panel of Figure~\ref{fig:cen_sat_agn}, we show the
same relation again, but using different symbols/colors to
differentiate the subsets of Seyfert galaxies, LINERs (low ionization
nuclear emission regions), LIRGs and merging galaxies. The
identification of Seyferts and LINERs is done using the Baldwin,
Phillips \& Terlevich (BPT) diagram \citep{Baldwin1981}, adopting the
AGN classification curve of \citet{Kauffmann2003} and the
Seyfert-LINER dividing line of \citet{Cid2010}. The relevant
emission-line ratios are taken from  the Max-Planck-Institute for
Astrophysics-John Hopkins University (MPA-JHU) SDSS database
\citep{Brinchmann2004}. The LIRGs in our sample are identified
according to their infrared luminosity over 8--1000 \micron,
$L_{8-1000\mbox{\micron}}$. To this end, we first estimate the
infrared luminosity over 40--500 \micron, $L_{40-500\mbox{\micron}}$,
based on the luminosities at 60 and 100 \micron\ from the
\textit{Infrared Astronomical Satellite} (\textit{IRAS}) database
\citep{Moshir1992} according to \citet[][see their Table
  1]{Sanders1996},  and then convert $L_{40-500\mbox{\micron}}$ to
$L_{8-1000\mbox{\micron}}$  adopting a conversion factor of 1.75
following \citet{Hopkins2003}.  The identification of interacting
galaxy systems is similar to  \citet{Lin2004}. In short, two close
galaxies are classified as a pair and included in the subset of
interacting systems if their projected separation is \textless\ 50
$h^{-1}$ kpc and their line-of-sight velocity difference is
\textless\ 500 km s$^{-1}$.  Additional interacting systems are
identified by visually inspecting the SDSS images.  We again compare
their distributions of residuals with that of the total sample,  and
K$-$S test returns a p-value = 0.23, 0.21, 0.66 and 0.24  for LINERs,
Seyfert, LIRGs and mergers, respectively.  Again, there is no/weak
differences between different types of galaxies in the right panel of
Figure~\ref{fig:cen_sat_agn}.  

These results demonstrate that Eq.~\ref{co_estim} provide unbiased
estimator of the CO luminosity of a galaxy, and can thus be applied to
large samples of galaxies for which CO observations are not available.

\begin{table*}
        \centering
	\caption{Best-fit relations for three-parameter molecular gas mass estimators.}
	\label{tab:tbl4}
	\begin{tabular}{cccccccc}
           
	\hline\hline 
$x_2$ parameter	&	$x_3$ parameter	&	$k_1$			&	$k_2$			&	$k_3$			&	b			&	$\sigma_{\rm int}$	& corresponding panel	\\\hline 
$g-r$	& log $({M_\ast}/{\rm M_\odot})$	&	0.77	$\pm$	0.03	&	0.29	$\pm$	0.08	&	0.28	$\pm$	0.04	&	-1.40	$\pm$	0.24	&	$\dots$	& Figure~\ref{fig:cen_sat_agn}\\
$g-r$	& log $({L_{r}}/{\rm L_\odot})$	&	0.82	$\pm$	0.03	&	0.55	$\pm$	0.06	&	0.18	$\pm$	0.03	&	-0.94	$\pm$	0.21	&	0.16	& the left panel of Figure~\ref{fig:oth_est} 	\\
$g-r$	& log $({L_{z}}/{\rm L_\odot})$	&	0.81	$\pm$	0.03	&	0.47	$\pm$	0.06	&	0.20	$\pm$	0.03	&	-0.99	$\pm$	0.20	&	0.16	& the middle panel of Figure~\ref{fig:oth_est} 	\\
log $({L_{r}}/{\rm L_\odot})$	& log $({L_{z}}/{\rm L_\odot})$	&	0.79	$\pm$	0.03	&	-1.89	$\pm$	0.22	&	2.07	$\pm$	0.21	&	-0.72	$\pm$	0.21	&	0.15 	& the right panel of Figure~\ref{fig:oth_est}	\\
 \hline
	\end{tabular}
	\begin{flushleft}
	\textbf{Notes.} 
	{The relations are parametrized as log~$(L_{\rm CO(1-0)}/({\rm K\;km\;s}^{-1}\;{\rm pc^{2}})) = k_1 log~(L_{\mbox{12\micron}}/{\rm L_{\odot}}) + k_2 x_2 + k_3 x_3 + b $, with all the quantities given in this table. The derived intrinsic scatter of each relation is listed as $\sigma_{\rm int}$.
 }
         \end{flushleft}
\end{table*}

\subsection{Alternative estimators}
\label{sec:alter}

It should be noted that the estimators mentioned in
Section~\ref{sec:3para} can only be used with \mstar\ measured in the
same manner as that used here, i.e. NSA \mstar, as there are
significant systematic uncertainties/biases on \mstar\ associated with
the different methods and assumptions used to infer it \citep[e.g.
  stellar initial mass function IMF and population models;
  e.g.][]{Bell-2003, Li-White-2009}.  In addition, for these
estimators, we can not separate the intrinsic scatter from the
total/observed scatter (as shown in Table~\ref{tab:tbl4}), as the NSA
catalogue does not include a random/measurement uncertainty on
\mstar\ for each galaxy.  According to the analysis of different
stellar masses estimated in different manner based on similar IMF, the
typical uncertainty is about 0.15 - 0.2 dex, but it increases
substantially at lower masses, reaching 0.3 dex at $10^8$ $h^{-2}$
\msolar.  We therefore provide other estimators using instead the
$r$-- and/or $z$-- band stellar luminosities in Table~\ref{tab:tbl4},
that are not subject to these large systematic effects and are less
dependent on stellar population models. 

These relations, shown in Figure~\ref{fig:oth_est}, arise naturally
from the data  through multiple-parameter linear regression fitting,
without any assumption.  All of these estimators appear to be
similarly good, with small total/observed scatters of $\sim$ 0.18
dex (intrinsic scatters of $\sim$ 0.16 dex) and no obvious
difference between central and satellite galaxies.

\subsection{The CO(2-1)-to-CO(1-0) line ratio $R21$}
\label{sec:est_R}

Here we present an immediate application of the $L_{\rm CO(1-0)}$
estimator (Eq.~\ref{co_estim}), whereby we estimate the CO(1-0)
luminosity of galaxies that have only CO(2-1) observations and then
investigate the inferred CO(2-1)-to-CO(1-0) line ratios ($R21$). For
this purpose we have compiled a sample of galaxies with CO(2-1) but no
CO(1-0) observations, including  72  galaxies from JINGLE
\citep{Saintonge2018},  27 from the SMT sample of \citet{Jiang2015},
and 10 from our own sample (see Table~\ref{tab:tbl2}).  We first
estimate the CO(1-0) luminosity of each galaxy from its
12\micron\ luminosity, stellar mass and $g-r$ color according to
Eq.~\ref{co_estim}, and then infer $R21$ from the ratio of the
observed CO(2-1) luminosity to the estimated CO(1-0) luminosity.  For
comparison, we also consider two samples of galaxies from previous
studies where both CO(1-0) and CO(2-1) integrated fluxes are
available:18 galaxies from the HERA CO-Line Extragalactic
  Survey \citep[HERACLES;][]{Leroy2009}, and 27 xCOLD GASS galaxies
  with detections in both IRAM CO(1-0) and APEX CO(2-1) measurements
\citep{Saintonge2017}. 

\begin{figure}
  \centering   \includegraphics[width=0.48\textwidth]{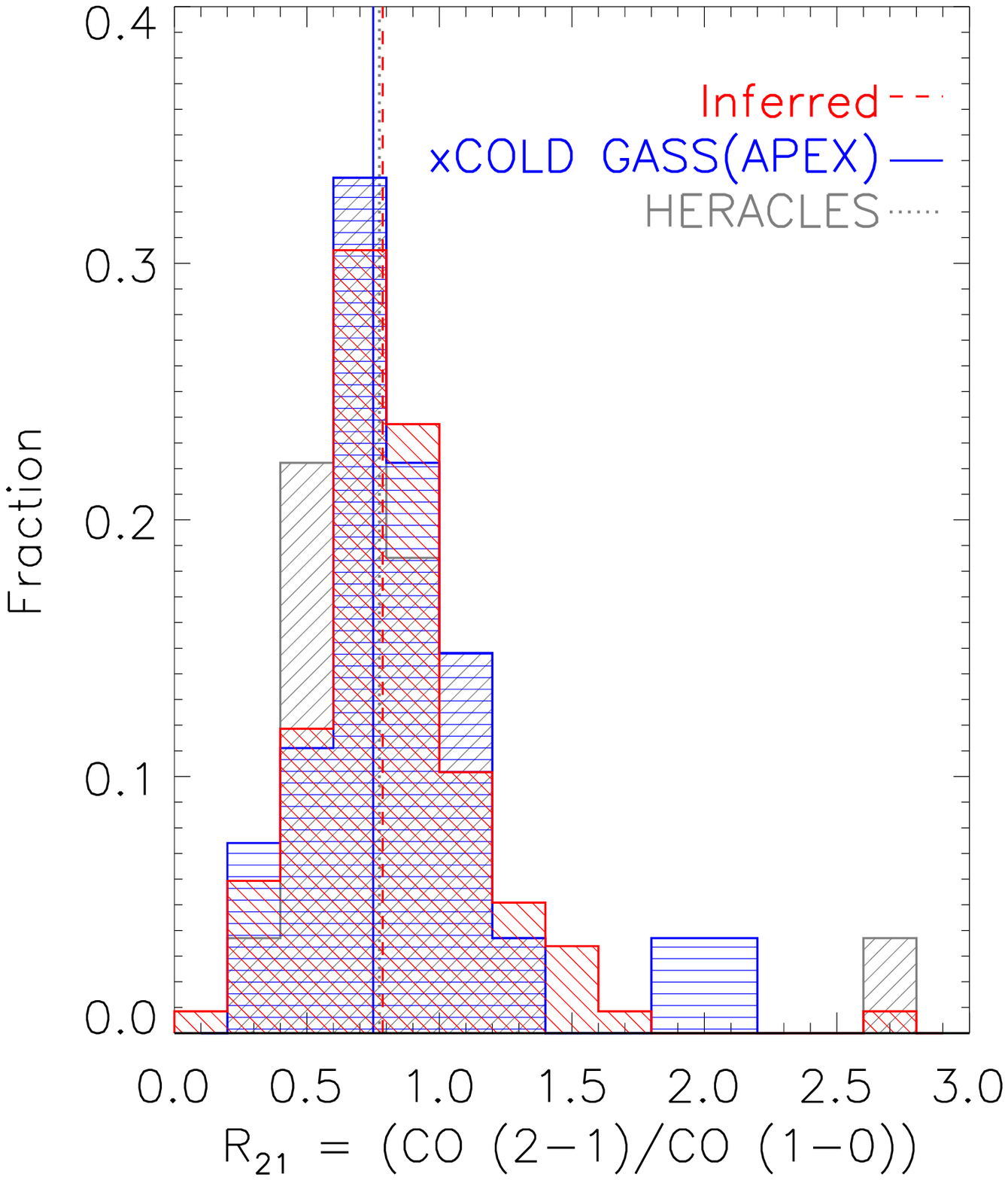} 
  \caption{ Distribution of CO(2-1)-to-CO(1-0) line ratios for the inferred (based on observed $L_{\rm CO(2-1)}$ 
    and predicted $L_{\rm CO(1-0)}$) and observed (literature) samples. The
    red histogram shows the inferred $R21$, 
    while the blue and gray histograms show the observed 
    xCOLD GASS (with APEX) and HERACLES survey, respectively.  
    The colored vertical lines indicate the median $R21$ of each sample. }
  \label{fig:hist_r21}
\end{figure}

Figure~\ref{fig:hist_r21} shows the histogram of the {\em inferred}
$R21$ (red), compared to those of the {\em observed} $R21$ from the
two previous studies considered (blue and gray). It is encouraging to
see that the $R21$ distribution based on our estimates agrees very
well with the real observations, that are also in good agreement with
each other. This is true in terms of both the median $R21$ of the
samples and the overall shape of the distributions. The median $R21$
is 0.78 for the {\em inferred} sample, very close to the median value
of 0.75 and 0.77 for xCOLD GASS and HERACLES. The $1\sigma$ widths of
the distributions are also very similar, 0.17, 0.18 and 0.18 dex,
respectively. The K$-$S test yields p-value of 0.96 and 0.76 for xCOLD
GASS and HERACLES, respectively, suggesting the probabilities are
larger than 75\% that they are drawn from the same sample.  Finally,
we nevertheless notice a tail of galaxies with higher-than-average
$R21$ in both xCOLD GASS and HERACLES, dominated by merging
systems. The only such galaxy in our inferred sample is also a merging
galaxy.  Given the relatively small sample sizes, however, this
different fraction of higher-than-average $R21$ should not be
overemphasized.

\begin{figure*}
\begin{center}
  \includegraphics[width=0.46\textwidth,clip=true]{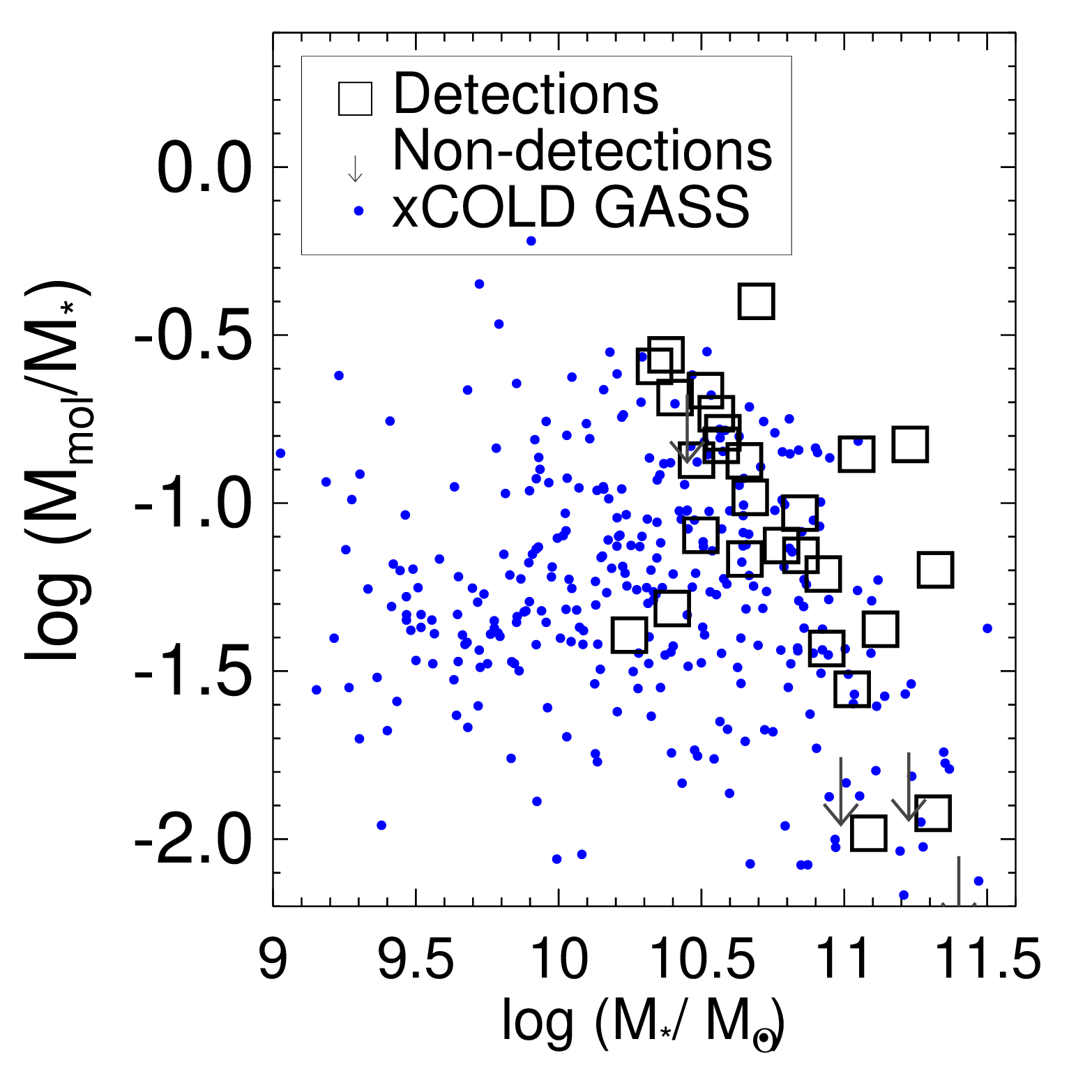}
  \includegraphics[width=0.46\textwidth,clip=true]{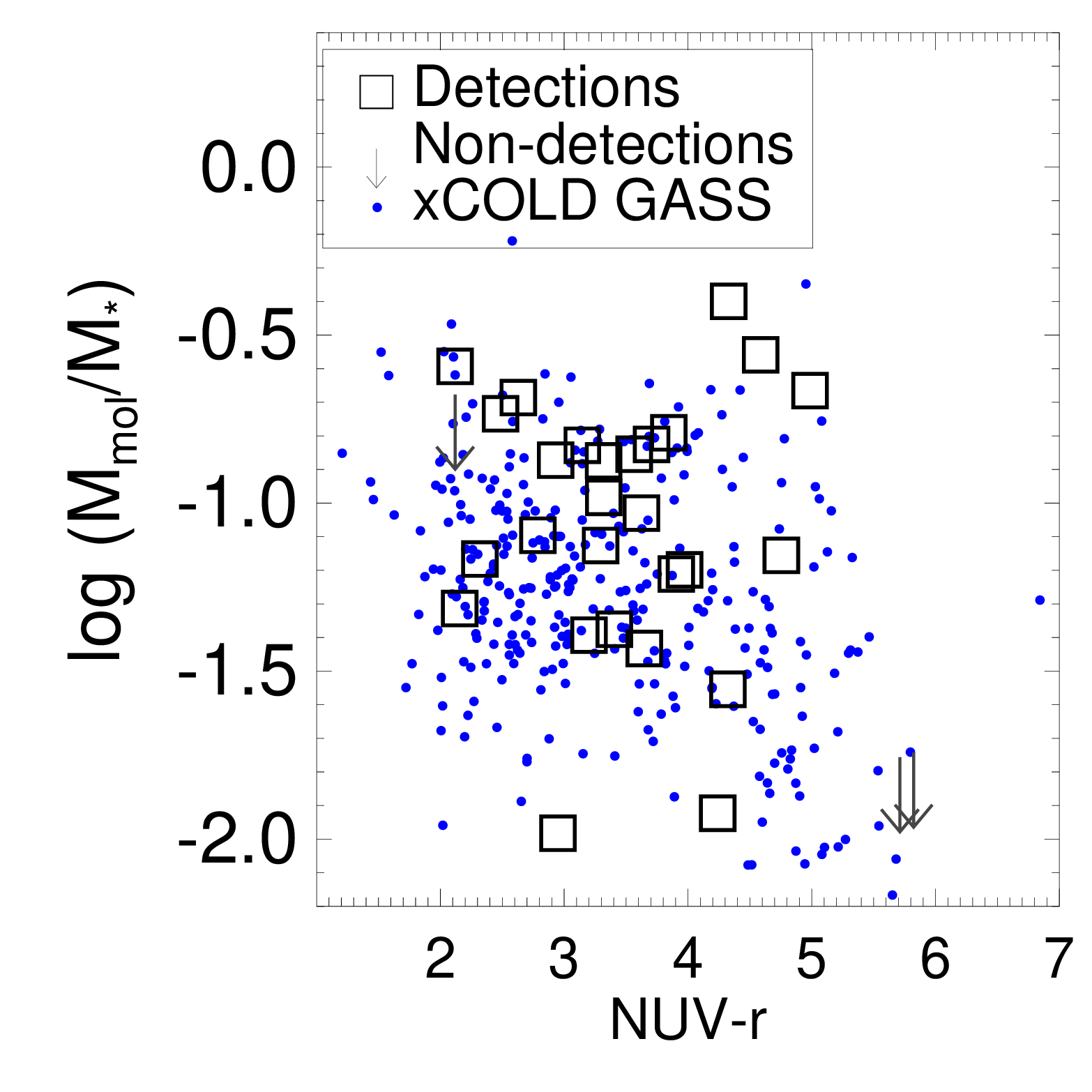}
  \caption{Distribution of our sample galaxies on the planes of
    molecular-to-stellar mass ratio log $(M_{\rm mol}/M_\ast)$ versus
    log $(M_\ast)$ (middle) and log $(M_{\rm mol}/M_\ast)$
    versus {\nuvr} (right). In each panel, our galaxies are plotted
    as large symbols (black squares for CO detections and downward-pointing arrows for upper limits), 
    while the detections from the xCOLD GASS survey are plotted as blue dots.}
\label{fig:sample_properties}
\end{center}
\end{figure*}

\section{Total molecular gas content \mh}
\label{sec:mol_gas}

In this section we examine the correlation of molecular gas mass
  fraction with star formation rate, using both \textit{real} gas masses
  from our sample and the xCOLD GASS and \textit{estimated} gas masses
  for the current MaNGA sample obtained from the three-parameter CO (1-0)
  luminosity estimator as presented
  in Section~\ref{sec:3para}.

\subsection{\mh\ of our sample}
\label{sec:comp_mh}

For the galaxies observed with the PMO 13.7-m telescope, we
  estimate the total molecular gas mass by multiplying the
  CO (1-0) luminosity (corrected for aperture effect)
  by a CO-to-H$_2$ conversion factor: $M_{\rm mol} = \alpha
  _{\rm CO}L_{\rm CO(1-0)}$. A Galactic conversion factor $\alpha_{\rm
    CO} = 4.3$ \msolar (K km s$^{-1} {\rm pc^{2}})^{-1}$ corresponding
  to $X_{\rm CO} = 2\times 10^{20}$ cm$^{-2}$ (K km s$^{-1})^{-1}$
  is adopted, so the resulting molecular gas mass includes a factor of
  1.36 for the presence of heavy elements (mainly helium). For the
  galaxies observed with JCMT or CSO, we convert their CO (2-1)
  luminosities to CO (1-0) luminosities assuming a CO (2-1)/CO (1-0) line
  ratio of $R21=0.7$ following \citet{Leroy2013}, and we caculate the
  total molecular gas mass in the same manner as above. Although
  slightly smaller than the median values of the samples studied in
  the previous section (Section~\ref{sec:est_R}), the line ratio of
  $R21=0.7$ is adopted for the convenience of potential comparisons of
  our results with the literature. The H$_2$ masses are listed
  in the last column of Table~\ref{tab:tbl2}.

Figure~\ref{fig:sample_properties} shows the molecular-to-stellar mass
ratio \mhms\ as a function of respectively stellar mass \mstar\ and
\nuvr\ color, this for both our galaxies and the  xCOLD GASS
detections. Although biased to massive gas-rich galaxies, our sample
spans wide ranges in both \nuvr\ and \mhms, similarly to the xCOLD
GASS sample galaxies, which allows the statistical analyses presented
below.

\subsection{Correlation of $M_{mol}$ with star formation rate}

\begin{figure*}
\begin{center}
\includegraphics[width=0.45\textwidth]{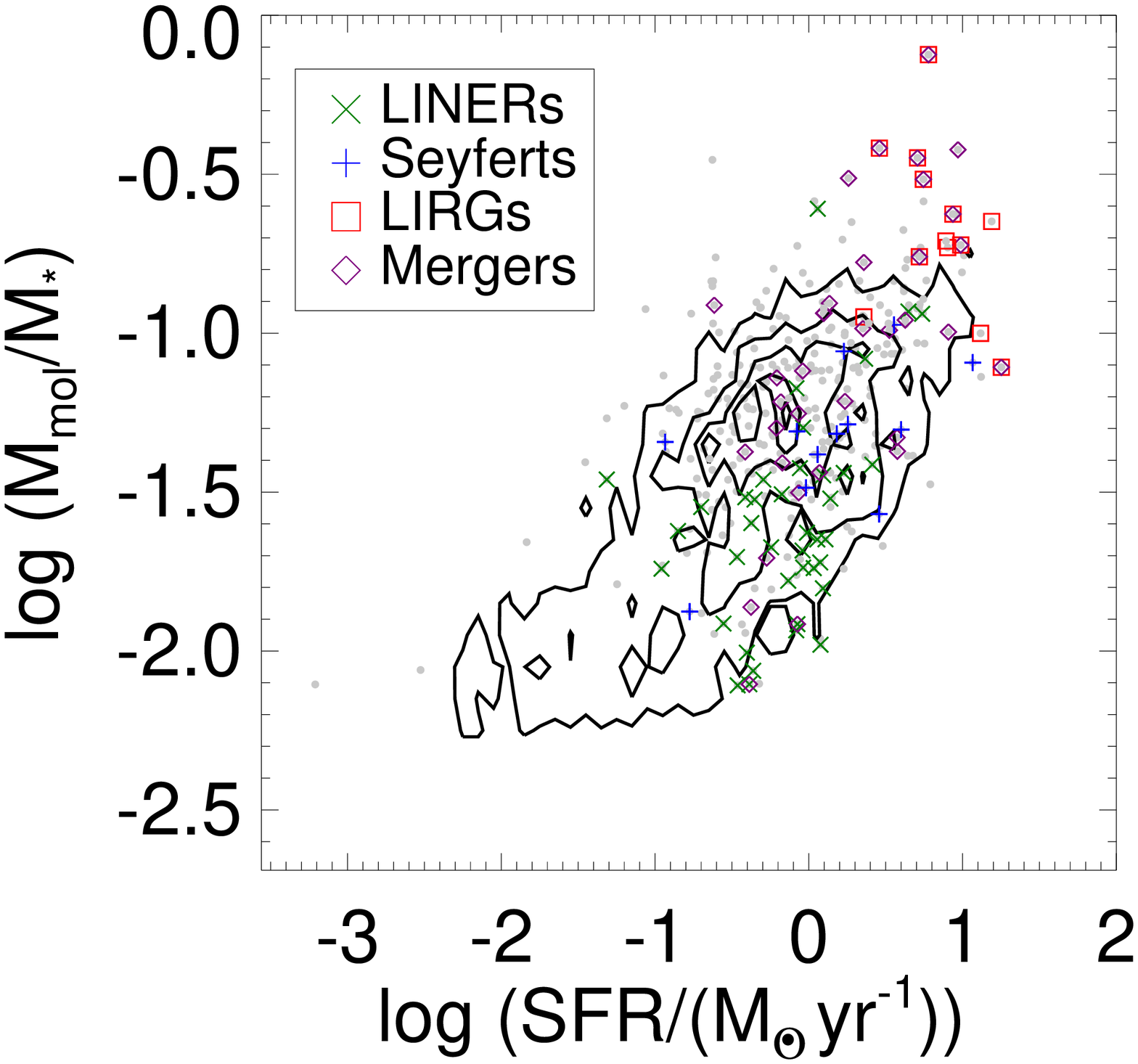}
\includegraphics[width=0.45\textwidth]{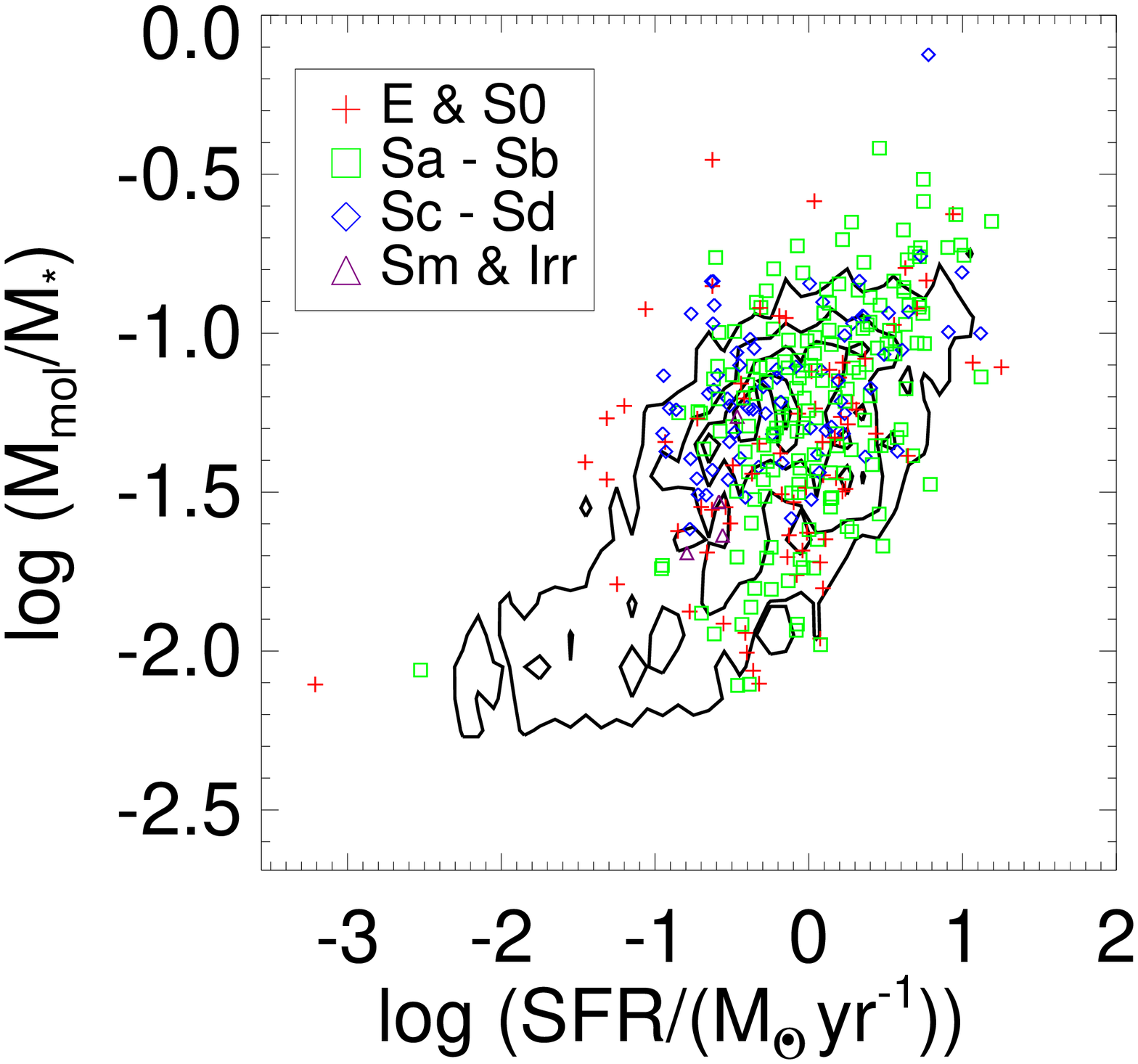}
\end{center}
\begin{center}
\includegraphics[width=0.45\textwidth]{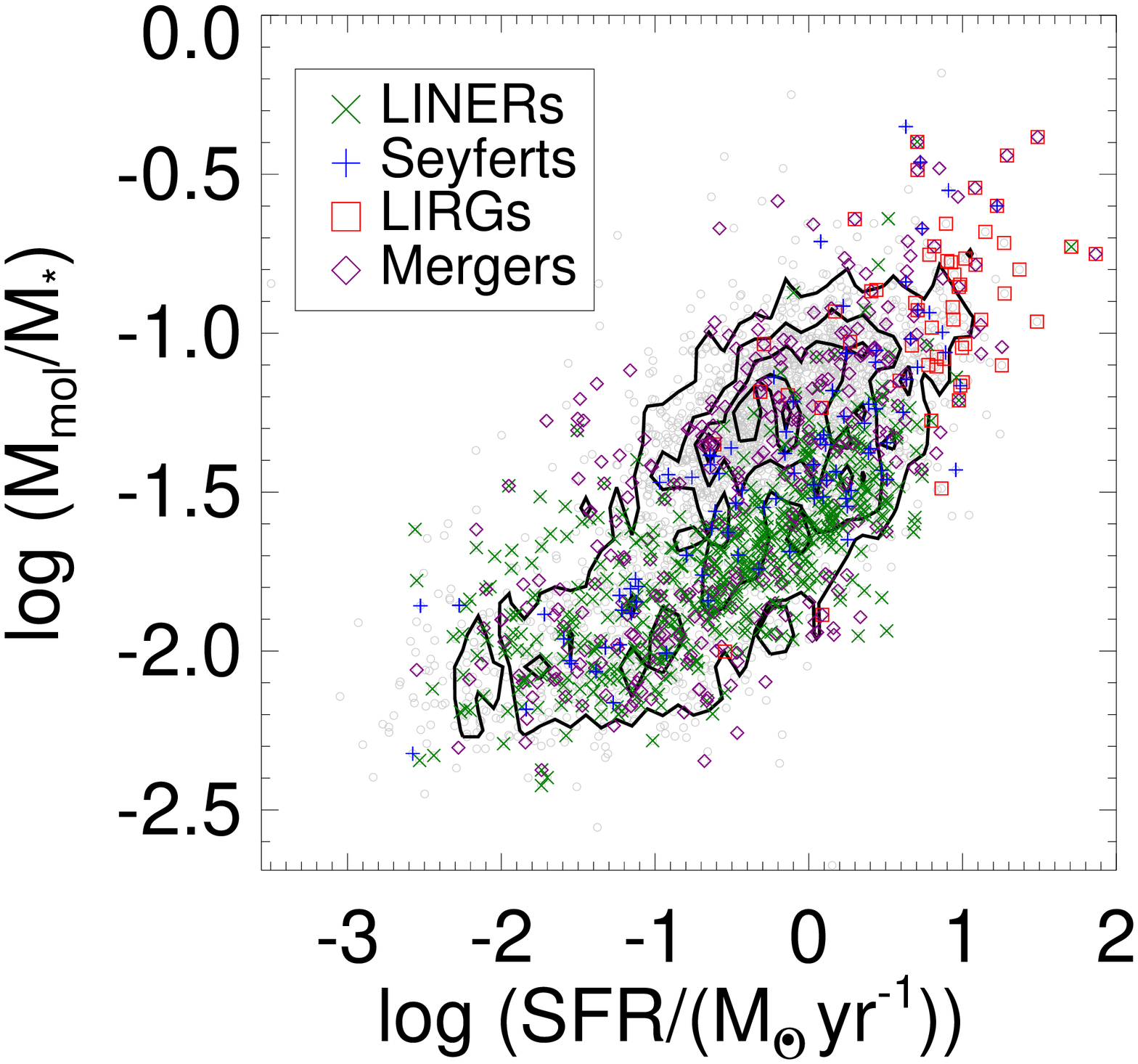}
\includegraphics[width=0.45\textwidth]{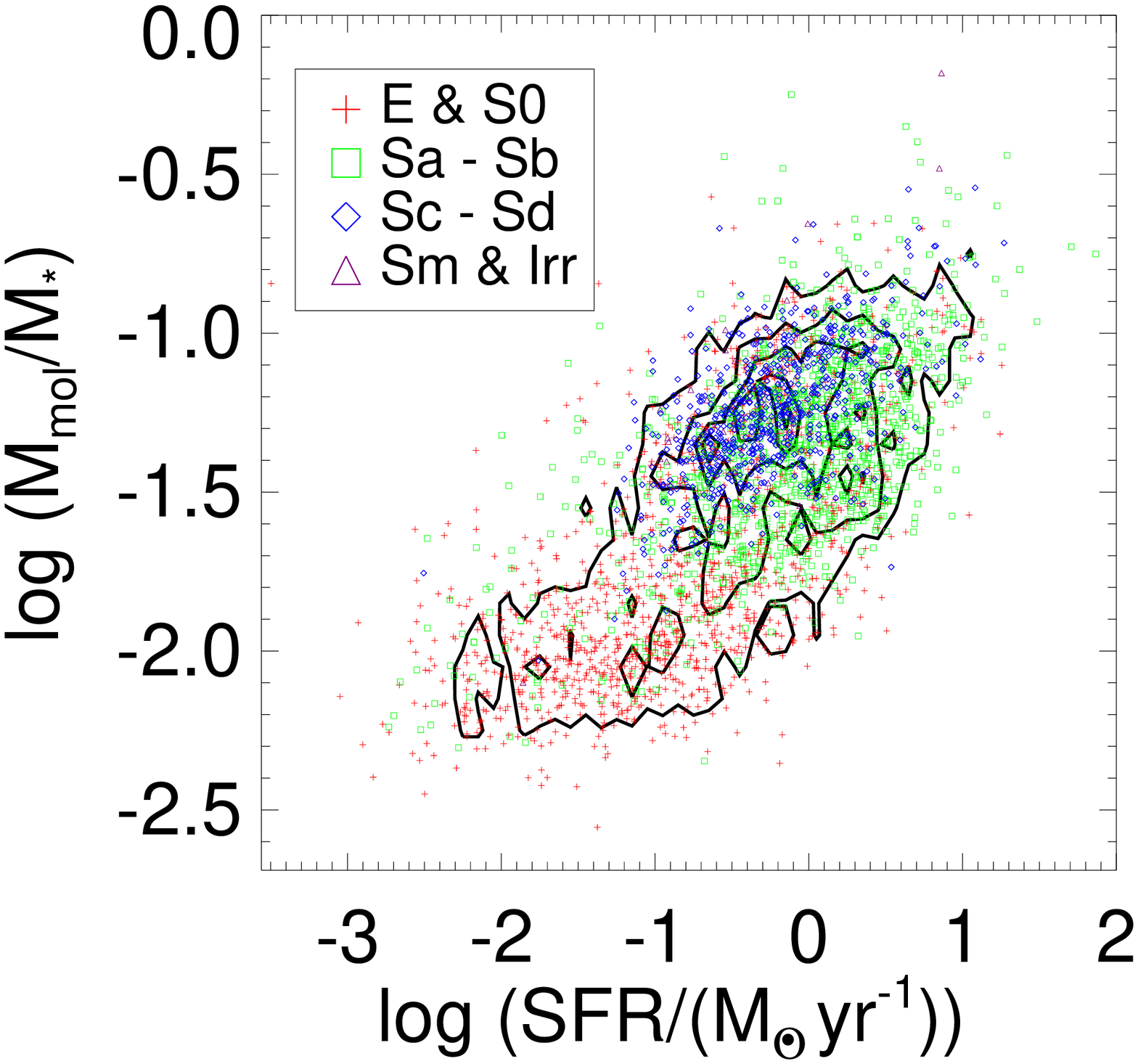}
\end{center}
\caption{ Distribution of galaxies on the plane of molecular
    gas-to-stellar mass ratio versus star formation rate. The upper
    panels show the distribution of our sample and xCOLD GASS with
    \textit{observed} gas masses, and the lower panels show the
    distribution of the MaNGA/MPL-8 sample for which the gas masses
    are \textit{estimated} using our three-parameter $L_{\mbox{CO(1-0)}}$
    estimator (Eq.~\ref{co_estim}). Different colors/symbols
    represent different types of galaxies (left) or different
    morphologies (right), as indicated. Black contours show the
    distribution of the whole MaNGA/MPL-8 sample, repeated in
    every panel for comparison.}
\label{fig:dist_sfr_mh}
\end{figure*}

Hereby we persent an application of our CO estimator to the
  current sample of the MaNGA survey, the MaNGA/MPL-8 sample which
  consists of 6,487 galaxies. In most cases CO observation is not
  available.  In particular, we focus on the correlation of molecular
  gas-to-stellar mass ratio, \mhms\ with the star formation rate
  (SFR). We estimate an H$_2$ mass for each galaxy in MaNGA/MPL-8
  using our three-parameter $L_{\mbox{CO(1-0)}}$ estimator (see
  Eq.~\ref{co_estim}), and we make use of the $M_\ast$ and SFR
  estimates from the \textit{GALEX}-SDSS-\textit{WISE} Legacy Catalog
  \citep[GSWLC;][]{Salim2016,Salim2018} for this analysis.  As in
  Section~\ref{sec:3para}, the galaxies are divided into subsets of
  LINERs, Seyfert galaxies, LIRGs and merging galaxies, as well as
  subsets of different morphological types according to Hubble type
  from the HyperLeda database \citep{Makarov2014}.  For comparison we
  also include in the analysis the xCOLD GASS and our sample which
  have \textit{observed} molecular gas masses.

In Figure~\ref{fig:dist_sfr_mh} we show the distributions of
  different types of galaxies in the \mhms---SFR plane.  The
  top panels display the galaxies from the xCOLD GASS and our sample,
  but highlighting subsets of LINERs/Syeferts/LIRGs/Mergers with
  different symbols/colors in the left panel and subsets of
  different morphologies in the right panel. The lower panels
  display the MaNGA/MPL-8 sample, highlighting the different subsets
  in the same manner as the upper panels. The black contours
  present the distribution of all MPL-8 galaxies, and are repeated
  in every panel for comparison.

We see that, overall, \mhms\ is positively correlated with SFR
  in all panels, as expected. When comparing the samples with
  \textit{real} gas mass fractions with the MaNGA/MPL-8 sample with
  \textit{estimated} gas mass fractions, we find similar correlations
  and scatters in the upper-right part of the diagram
  ($\log(M_{\mbox{mol}}/M_\ast)\ga -2$ and $\log SFR \ga -1$) where
  data is available for both samples. This is true not only for the
  whole samples, but also for subsamples of different types. 
  LIRGs are expectably located in the upper-right corner with highest SFRs and
  molecular gas fractions. Other types of galaxies span a wide range
  in both SFR and $M_{\mbox{mol}}/M_\ast$, but present systematic
  differences. For instance, at given SFR, merging galaxies appear to
  be more gas-rich than LINERs, while Seyfert galaxies are found in
  between; on average Sc and Sd-type spirals are more gas-rich than
  earlier morphological types which are distributed more broadly and
  scatteringly. These trends are more clearly seen in the lower panels
  thanks to the much larger sample sizes.
  
A remarkable difference between the real gas samples and the
  estimated gas samples occurs in the lower-left part of the panels
  where the MaNGA sample extends well below the detection limit of the
  xCOLD GASS, thus adding a significant population of gas-poor galaxies
  to this diagram which are predominantly early-type (E or S0).
  In contrast to the gas-rich galaxies from xCOLD GASS, the gas-poor
  population presents almost no correlation between the molecular gas mass
  fraction and the SFR, although their SFR spands a wide range from
  $\log(SFR/M_\odot{\mbox{yr}}^{-1})\sim-1$ to $\sim-3$. The
  lower-left panel shows that the majority of these galaxies are
  LINERs, while many mergers and some Seyferts also fall in this
  regime. The mergers should be \textit{dry mergers} given their
  relatively low gas fractions. All these trends are interesting,
  and deserve more detailed analyses. In the next work we will come
  back to the MaNGA sample, and we will combine
  the estimated global molecular gas mass with integral field
  spectroscopy to better understand the role of cold gas in
  driving the star formation and nuclear activity of galaxies.

\section{Summary and Discussion}
\label{sec:summary}

We have obtained CO(1-0) and/or CO(2-1) spectra for a sample of 31
galaxies selected from the ongoing MaNGA survey.  We  utilized three
different telescopes: PMO 13.7-m millimeter telescope located in
Delingha, China, CSO and JCMT located in Hawaii. We measured the total
CO flux/luminosities and molecular gas masses of our
galaxies. Combining our sample with other samples of CO observations
from the literature, we examined the correlations of the CO
luminosities with the infrared luminosities at 12
($L_{\mbox{12\micron}}$) and 22 \micron\ ($L_{\mbox{22\micron}}$). We
then examined the residuals of the $L_{\rm
  CO(1-0)}$-$L_{\mbox{12\micron}}$ relation as a function of  a
variety of galaxy properties including stellar mass ($M_\ast$), color
(\nuvr\ and $g-r$) and Sersic index ($n$), to find a linear
combination of multiple parameters that may be used to estimate
molecular gas masses for large samples of galaxies.We applied the
resulting best-fitting estimator to a sample of galaxies with only  CO
(2-1) observations and investigated the resulting CO(2-1)-to-CO(1-0)
ratios, as well as the current sample of MaNGA to study
  the correlation of molecular gas-to-stellar mass ratio as a function
  of SFR, galaxy type and morphology.

Our main conclusions can be summarized as follows:

\begin{enumerate}

\item Our sample consists of 31 relatively massive galaxies with
  stellar masses $\gtrsim 2\times10^{10}$ M$_\odot$, spanning all
  morphological types and covering wide ranges of colors and  molecular
  gas-to-stellar mass ratios, that are similar to those of the xCOLD GASS sample.

\item The CO luminosities are tightly correlated with the MIR
  luminosities, and the correlation with the 12 \micron\ band has a smaller 
  scatter and is  more linear than the one with the 22 \micron\ band.  The
  $L_{\rm CO(1-0)}$--$L_{\mbox{12\micron}}$ relation shows no/weak
  dependence on the MIR luminosities and Sersic indices, while galaxies with
  the lowest masses and/or bluest colors are below the mean relation.
  
\item A linear combination of the 12 \micron-band luminosity
  $L_{\mbox{12\micron}}$, stellar mass ($M_\ast$) and optical
  color $g-r$  provides unbiased estimates of the CO(1-0) luminosities (and
  thus  molecular gas masses). This estimator works well for
  both central and satellite galaxies, and for different types of
  galaxies such as LIRGs, Seyferts, LINERs and mergers. Replacing $M_\ast$ by the
  luminosity in the $r$ or $z$ band yields similarly good
  estimators.

\item The distribution of $R21$ obtained from estimated CO(1-0)
  luminosities agrees well with real measurements from previous studies.
  The median $R21$ of this inferred sample is $R21=0.78$,  with a
  scatter of 0.17 dex, consistent with the typical value
  of 0.7 that is commonly-adopted in previous studies.
  
\item Applying our $L_{\mbox{CO(1-0)}}$ estimator to
  $\sim6,400$ galaxies in the current sample of MaNGA, we find
  a significant population of gas-poor galaxies which are
  predominantly early-type. The molecular gas-to-stellar mass
  ratio of these galaxies shows no correlation with star formation
  rate, in contrast to gas-rich galaxies that have been previously studied in depth.

\end{enumerate}

The tight correlation between CO luminosity and  12 {\micron}
luminosity was already reported and studied in some detail by
\citet{Jiang2015}. This correlation is not unexpected, given both the
tight relation between the surface densities of star formation rate 
and cold gas mass, known as the Kennicutt-Schmidt law
\citep{Schmidt1959, Kennicutt1998}, and the correlations of SFR with
infrared luminosities. For the latter,  in particular,
\citet{Donoso2012} found that $\sim$ 80\% of the 12 \micron\ emission from
star-forming galaxies in the 	\textit{WISE} survey is produced by stellar
populations younger than $\sim$ 0.6~Gyr, implying a strong correlation of SFR
with the 12 \micron\ luminosity.

When compared to $L_{\mbox{12\micron}}$, the luminosities at 22\micron\ show 
larger scatter and a more non-linear correlation with the CO
luminosities. This is true for both CO(1-0) and CO(2-1) (see
Figures~\ref{fig:co10_w} and~\ref{fig:co21_w}). Assuming the SFR is the
driving factor for the correlations of $L_{\rm CO}$ with MIR
luminosities, this result implies that the 12 \micron\ luminosity is a
better indicator of star formation than the 22 \micron\ luminosity.  
This may be understood from the fact that the W3 band of \textit{WISE}, which spans
a wavelength range from 7.5 to 16.5 \micron, includes and is almost
centered on the prominent polycyclic aromatic hydrocarbon (PAH)
emission at 11.3 {\micron} \citep{Wright2010}, and that our sample does not 
include any strong AGN (e.g. quasar).
The mid-infrared spectra from
the Spitzer Infrared Nearby Galaxies Survey (SINGS; \citealp{Kennicutt2003}),
as obtained by the Spitzer Infrared Spectrograph \citep{Houck2004},
revealed that the PAH emission can contribute up to $\sim 20\%$ of the total infrared emission. 
In addition, both the dust continuum emission and the PAH emission 
are well correlated with CO emission \citep{Wilson2000, Cortzen2019}, 
while the W4 band of \textit{WISE} may detect stochastic emission from heated small grains 
with temperatures of $\sim\ 100-150$ K \citep[in addition to the Wien tail of thermal 
emission from large grains;][]{Wright2010}.

The $L_{\rm CO(1-0)}$--$L_{\mbox{12\micron}}$ relation should in
principle provide a very useful estimator, that can be easily applied
to estimate the molecular gas masses of large samples of galaxies,
particularly considering the all-sky survey data from 	\textit{WISE}
and the small scatter ($\sim$ 0.2 dex) of the relation. In fact, this
simple estimator has been successfully applied by the JINGLE team to
estimate observing times for the purpose of target selection
\citep{Saintonge2018}. In this work, we have further improved the
estimator by including two more parameters, $M_\ast$ and $g-r$, that
are also available for large samples of galaxies thanks to the imaging
data from SDSS and other large optical surveys. We have shown that
such estimators provide unbiased CO luminosity estimates for different
types of galaxies. Our new three-parameter estimator will be helpful
to provide more accurate estimates of molecular gas masses and thus to
study gas-related processes in a wider range of galaxies than
currently possible \citep[e.g. gas-poor galaxies and gas-related
  quenching processes; e.g.][]{Li2012, Zhang2013}. As an
  example, in this paper we have performed a quick application of our
  estimator to the current MaNGA sample (MPL-8), and found a
  significant population of gas-poor galaxies that fall below the
  detection limit of existing CO surveys (e.g. xCOLD GASS). This
  population is dominated by early-type galaxies and shows no
  correlation between $M_{\mbox{mol}}/M_\ast$ and SFR, differently
  from gas-rich galaxies which show a strong correlation.  We will
  come back to the MaNGA sample in future works and combine our
  estimated gas masses with the MaNGA integral field spectroscopy to
  better understand this gas-poor population.

\section*{Acknowledgements}
We are grateful to the anonymous referee for his/her
  detailed comments which have improved our paper.
We thank the staff at Qinghai Station of the PMO, and the
JCMT and CSO for continuous help with observations and data reduction.
This work is supported by the National Key R\&D Program of China
(grant Nos. 2018YFA0404502), the National Key Basic Research Program
of China (grant No. 2015CB857004) and the National Science Foundation
of China (grant Nos. 11821303, 11973030, 11761131004, 11761141012, and 11603075), and 
CDW acknowledges support from the Natural Science and Engineering 
Research Council of
Canada and the Canada Research Chairs program.

We are grateful to the MPA-JHU group for access to their
data products and catalogues. The Starlink software
\citep{Currie2014} is currently supported by the East
Asian Observatory.
This work has made use of data from the HyperLeda
  database (http://leda.univ-lyon1.fr).

The James Clerk Maxwell Telescope is operated by the East Asian
Observatory on behalf of The National Astronomical Observatory of
Japan; Academia Sinica Institute of Astronomy and Astrophysics; the
Korea Astronomy and Space Science Institute; Center for Astronomical
Mega-Science (as well as the National Key R\&D Program of China with
No. 2017YFA0402700). Additional funding support is provided by the
Science and Technology Facilities Council of the United Kingdom and
participating universities in the United Kingdom and Canada.
The authors wish to recognize and acknowledge the very
significant cultural role and reverence that the summit of Maunakea has always had within the indigenous Hawaiian community. We are most fortunate to have the opportunity to conduct observations from this mountain.

    Funding for the SDSS and SDSS-II has been provided by the Alfred P. Sloan Foundation, the Participating Institutions, the National Science Foundation, the U.S. Department of Energy, the National Aeronautics and Space Administration, the Japanese Monbukagakusho, the Max Planck Society, and the Higher Education Funding Council for England. The SDSS Web Site is http://www.sdss.org/.

    The SDSS is managed by the Astrophysical Research Consortium for the Participating Institutions. The Participating Institutions are the American Museum of Natural History, Astrophysical Institute Potsdam, University of Basel, University of Cambridge, Case Western Reserve University, University of Chicago, Drexel University, Fermilab, the Institute for Advanced Study, the Japan Participation Group, Johns Hopkins University, the Joint Institute for Nuclear Astrophysics, the Kavli Institute for Particle Astrophysics and Cosmology, the Korean Scientist Group, the Chinese Academy of Sciences (LAMOST), Los Alamos National Laboratory, the Max-Planck-Institute for Astronomy (MPIA), the Max-Planck-Institute for Astrophysics (MPA), New Mexico State University, Ohio State University, University of Pittsburgh, University of Portsmouth, Princeton University, the United States Naval Observatory, and the University of Washington.


\begin{appendix}

\section{The capability of PMO to detect external galaxies}
\label{appendix}
\begin{figure}
\begin{center}
\includegraphics[scale=0.34]{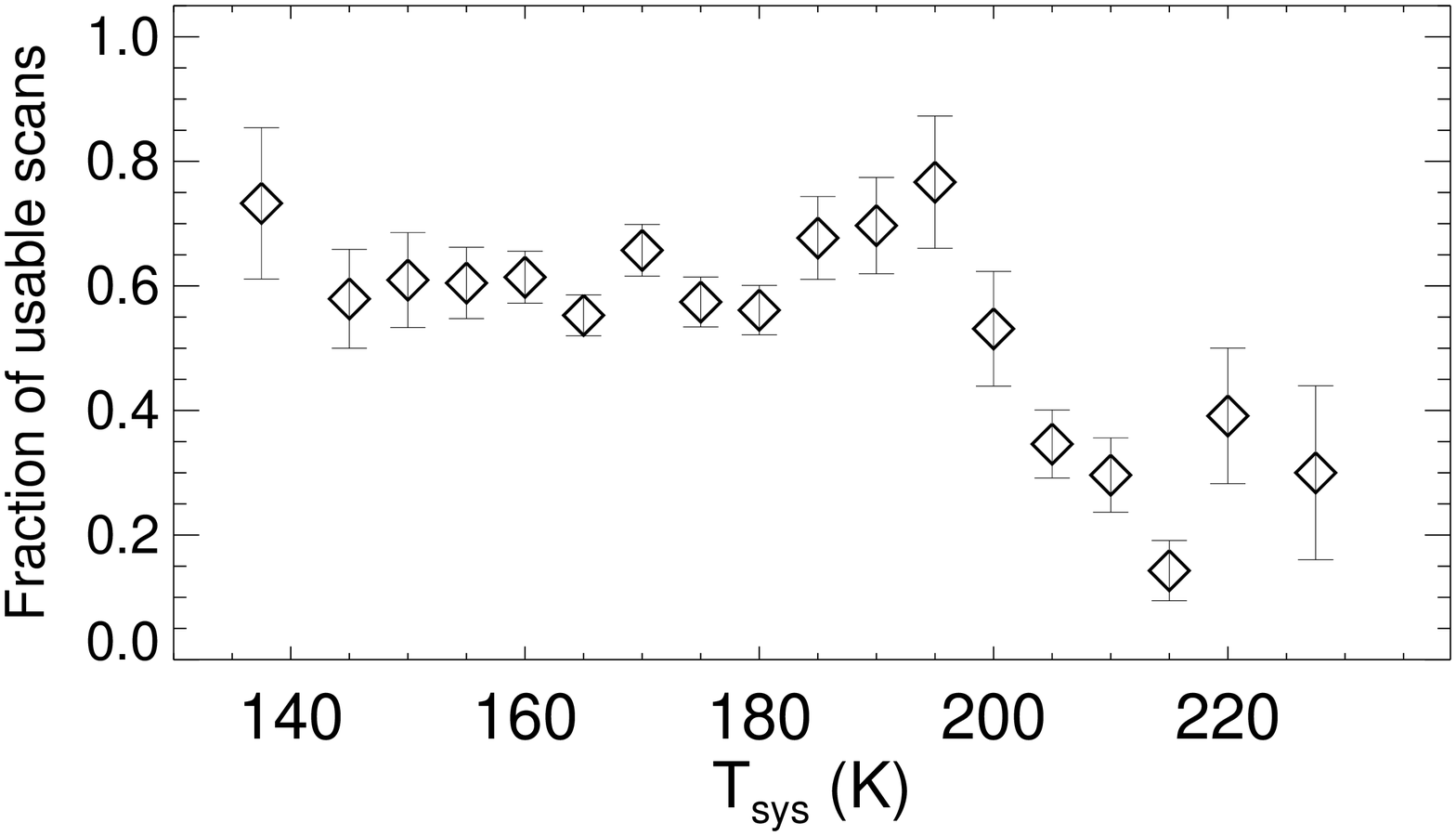}
\caption{ Fraction of usable scans taken with the PMO telescope as
  a function of the system temperature. The diamonds and  error bars
  are the mean and $1\sigma$ scatter of the fraction within each $T_{\rm sys}$ bin.}
\label{fig:t_sys}
\end{center}
\end{figure}

The PMO 13.7-m telescope has traditionally been mostly used to observe
Galactic sources \citep[e.g.][]{Ma2019} or extra-galactic sources in
the  very nearby Universe ($z<0.01$; e.g. \citealp{Li2015}). All our
targets are beyond $z=0.01$.  Our work is thus the first attempt to
observe a sample of non-local galaxies with this telescope. Therefore,
this is a good opportunity to test the capability of PMO to detect
external galaxies. We find the fraction of usable scans to strongly
depend on the system temperature, $T_{\rm sys}$. This is clearly shown
in Figure~\ref{fig:t_sys}, where we plot the fraction  of selected
scans as a function of $T_{\rm sys}$.  The fraction is roughly
constant at 60\% when $T_{\rm sys}\lesssim 200$ K,  but it decreases
dramatically at higher temperatures.  At $T_{\rm sys}>220$ K, the
observation efficiency is very low, with only $20-30\%$ of usable
scans.  Figure~\ref{fig:t_sys} shows that the PMO telescope can be
effectively used to observe external galaxies,  as long as the system
temperature is lower than $\sim 200$ K. Our observations were
carried out in two period, one in May and one in winter. Typically,
the system temperature in the winter period of our observations ranges
from 150 to 220 K.  In May, however, the situation is already much
worse, with a mean of $T_{\rm sys} \sim 200$ K.  Therefore, most of
the discarded scans were taken in May (after around mid-May).  In
total, the effective on-source time is $\sim$ 75 hours for the
observations in the winter period, but only 95 minutes for those in
May, although the  actual allocated time was much longer in the latter
period.

\end{appendix}
\bibliographystyle{aasjournal}
\bibliography{apj}

\end{document}